\documentstyle [12pt] {article}
\begin{document}
\setcounter{page}{1}
\def\theequation{\arabic{section}.\arabic{equation}}
\def\theequation{\thesection.\arabic{equation}}
\setcounter{section}{0}

\noindent Nuclear Physics {\bf A} (in press)

\vspace*{1.0cm}
\begin{center}
\section*{\bf On the relativistic \\field theory model of the deuteron II}
\smallskip

\vspace{30pt}
{\bf A. N. Ivanov$^1$, N.I. Troitskaya\footnote{\normalsize E--mail: ivanov@kph.tuwien.ac.at,Tel.:+43--1--58801--5598, Fax:+43--1--5864203\\ Permanent Address:
State Technical University, Department of Theoretical
Physics, 195251 St. Petersburg, Russian Federation.}, M. Faber\footnote{\normalsize E--mail: faber@kph.tuwien.ac.at, Tel.: +43--1--58801--5598, Fax: +43--1--5864203}, H. Oberhummer\footnote{\normalsize E--mail: ohu@kph.tuwien.ac.at, Tel.: +43--1--58801--5574, Fax: +43--1--5864203}}
\vskip1.0truecm

{\it Institut f\"ur Kernphysik, Technische Universit\"at Wien,}\\
\baselineskip=14pt
{\it Wiedner Hauptstr. 8-10, A-1040 Vienna, Austria}\\

\end{center}

\vskip1.0truecm
\begin{center}
\begin{abstract}
The relativistic field theory model of the deuteron suggested in [1] is revised and applied to the calculation of the cross sections of the low--energy radiative neutron--proton capture n + p $\to$ D + $\gamma$ and the low--energy two--proton fusion p + p $\to$ D + $e^{\,+}$ + $\nu_{\rm e}$. For the low--energy radiative neutron--proton capture n + p $\to$ D + $\gamma$ our result agrees well with both the experimental data and the potential model prediction. In the case of the two--proton fusion the cross section obtained is 2.9 times as much as that given by the potential approach. The obtained result is discussed in connection with the solar neutrino problem.
\end{abstract}
\end{center}
\vskip1.0truecm

\vskip1.0truecm
\begin{center}
PACS:11.10.Ef, 13.75. Cs, 14.20. Dh.\\
\noindent Keywords: relativistic field theory, deuteron, neutron, proton, radiative capture, weak interaction, W--boson, fusion
\end{center}
\vskip1.0truecm

\newpage

\section{Introduction}

\hspace{0.2in} In a recent publication [1] we have suggested a relativistic field theory model of the deuteron. This model, being some kind of a $\sigma$--model [2], is based on the assumption that the physical deuteron state should be produced due to integration over low--energy proton--neutron fluctuations at energies restricted by the scale $\Lambda_{\rm D}$. For simplification we have suggested a one--nucleon loop approximation for the integration over proton--neutron fluctuations. Unfortunately, this approximation does not have a perturbative parameter and should be considered as effective one, like that suggested by Nambu and Jona--Lasinio [3]. In this case the scale $\Lambda_{\rm D}$ has the meaning of the cut--off. We have defined $\Lambda_{\rm D}$ in terms of the effective radius of the deuteron $r_{\rm D}$, i.e. $\Lambda_{\rm D}\,=\,1/r_{\rm D}$. For the estimate of $\Lambda_{\rm D}$ we have applied the non--relativistic formula: $r_{\rm D}\,=\,(\varepsilon_{\rm D}\,M_{\rm N})^{-\,1/2}\,=\,4.319\,{\rm fm}$ [4,5], where $\varepsilon_{\rm D}\,=\,2.225\,{\rm MeV}$ is the binding energy of the physical deuteron [5] and $M_{\rm N}\,=\,938\,{\rm MeV}$ is the mass of the nucleon. We used equal masses for the proton and neutron, i.e. $M_{\,\rm p}\,=\,M_{\rm n}\,=\,M_{\rm N}\,=\,938\,{\rm MeV}$. This corresponds to the chiral limit when masses of current u-- and d--quarks vanish, i.e. $m_{\,0\,u}\,=\,m_{\,0\,d}\,=\,0$. Our estimate of $\Lambda_{\rm D}$ gives: $\Lambda_{\rm D}\,=\,1/r_{\rm D}\,=\,46\,{\rm MeV}$.

The interactions of the deuteron with proton and neutron were described in terms of two coupling constants $g_{\rm V}$ and $g_{\rm T}$, defining the interactions via vector and tensor nucleon currents, respectively. In the one--nucleon loop approximation we calculated the anomalous magnetic dipole moment of the deuteron $\kappa_{\rm D}$ in units of the nucleon magneton $\mu_{\rm N}\,=\,e/2\,M_{\rm N}$ , where $e$ is the proton charge, and  $Q_{\rm D}$ the electric quadrupole moment. Imposing the constraint $\kappa_{\rm D}\,=\,0$, being valid in the lowest approximation, we found the correlation between coupling constants $g_{\rm V}$ and $g_{\rm T}$, i.e. $g_{\rm T}/g_{\rm V}\,=\,-\,\sqrt{3/8}$. Then the coupling constant $g_{\rm V}$ has been fixed in terms of the electric quadrupole moment $Q_{\rm D}$. Eventually we calculated in one--nucleon loop approximation the binding energy of the physical deuteron in terms of $g_{\rm V}$ and $\Lambda_{\rm D}$. The theoretical result obtained was in good agreement with the experimental value: $(\varepsilon_{\rm D})_{\,\exp}\,=\,2.225\,{\rm MeV}$ [5].

The physical nature of the coupling constants $g_{\rm V}$ and $g_{\rm T}$ is connected with one--meson exchange. Of course, the $\pi$--meson exchange should give the main contribution. This assumption has been confirmed by the magnitude of $g_{\rm V}$, i.e. $g_{\rm V}\simeq g_{\,\pi\rm NN}$ , where $g_{\,\pi\rm NN}$ is the coupling constant of the ${\pi\rm NN}$--interaction.

The application of the model to the calculation of processes of low--energy interactions of the deuteron encounters the problem of the unambiguous computation of one--nucleon loop digrams that should describe the interactions of the deuteron with other particles in the suggested model. 

Indeed, since these are fermion loops, most of them, contributing to processes like radiative neutron--proton capture  n + p $\to$ D + $\gamma$, low--energy proton--proton reaction p + p $\to$ D + $e^{\,+}$ + $\nu_{\rm e}$ and so on, depend explicitely on the shift of virtual momenta of nucleons in the loop. This introduces substantial ambiguities interfering with the direct application of the model. However, free parameters appearing due to shifts of virtual momenta in nucleon loops must not be considered as free parameters of the model introduced especially for the computation of every nucleon diagram.

In this paper we revise our model [1] and apply it to the calculation of the cross sections of the radiative  neutron--proton capture  n + p $\to$ D + $\gamma$ and two--proton fusion p + p $\to$ D + $e^{\,+}$ + $\nu_{\rm e}$. In order to compute unambigously the effective anomalous magnetic and electric quadrupole moments of the deuteron we 
impose constraints on the computation of one--nucleon loops, caused by the requirement of electromagnetic gauge invariance of contributions for individual loops. This distinguishes our effective model from a field theory like QED, where the electromagnetic gauge invariance should be required for the complete set of diagrams in fixed order of perturbation theory. The requirement of electromagnetic gauge invariance applied to the individual nucleon loops allows one to fix ambiguities and use one--nucleon loops as well--defined quantum--field--theory objects. In order to remove ambigutities of the computation of the one--nucleon loop diagrams describing the effective Lagrangian of the radiative neutron--proton capture we turn to the selection rules. In the case of the computation of the effective Lagrangian responsible for the low--energy two--proton fusion p + p $\to$ D + $e^{\,+}$ + $\nu_{\rm e}$ we can remove ambigutities of the computation of the one--nucleon loop diagrams by using the gauge invariance under gauge transformations of the deuteron field.

The present paper is organized as follows. In Sect.~1 we adduce the starting assumptions that we have put in the foundation of the model, and phenomenological parameters in terms of the parameters characterizing the physical deuteron. In Sects.~2 and 3 we give a detailed derivation of the effective Corben--Schwinger and Aronson Lagrangians, respectively, describing electromagnetic interactions of the deuteron in terms the magnetic and electric quadrupole moments. In Sect.~4 and Sect.~5 we apply the relativistic field theory model of the deuteron to the computation of the amplitudes and cross sections of the radiative neutron--proton capture n + p $\to$ D + $\gamma$ and the low--energy proton--proton fusion p + p $\to$ D + $e^{\,+}$ + $\nu_{\rm e}$, respectively. In the Conclusion we discuss the obtained results.

\section{The deuteron structure in the one-nucleon loop approximation}
\setcounter{equation}{0}

\hspace{0.2in} In Ref. [1] we have suggested a relativistic field theory model for 
the  deuteron as a bound state of proton and neutron. The main idea, 
that has been put into the foundation of the approach, has been 
adopted from the Nambu--Jona--Lasinio (NJL) model [3]. As has been suggested by Nambu and Jona--Lasinio [3], the physical deuteron should appear as the proton--neutron collective excitation due to a local four--nucleon interaction and vacuum fluctuations of nucleons, calculated in one--nucleon loop approaximation, producing both a kinetic term of the physical deuteron and its interactions with other particles. Unfortunately, the direct realization of the Nambu--Jona--Lasinio program, i.e. the use of the local four--nucleon interaction, would lead to a strongly bound localized proton--neutron state, whereas a physical deuteron is weakly bound and rather "extended" object. Therefore, we have borrowed from the NJL model only the admission of the applicability of the one--loop approximation. This is used for the computation of observed parameters for physical states in  leading 
order in the long--wavelength expansion [6--9].

In our description of the physical deuteron, being some kind of the $\sigma$--model, we use the following 
scheme: We start from the Lagrangian of an unphysical deuteron field 
$D^{(0)}_\mu(x)$, considered as a bound proton--neutron state at 
zero binding energy and with a mass equal to the sum of the proton 
and neutron masses. Then in the one--nucleon loop approximation and 
leading order in the long--wavelength expansion we obtain an effective 
Lagrangian of a physical deuteron field describing a physical deuteron 
with observable binding energy $(\varepsilon_{\rm D})_{\rm exp}=2.225\,{\rm MeV}$ [5], 
anomalous magnetic dipole moment $(\kappa_{\rm D})_{\rm exp}=-0.023$, 
determined as $\kappa_{\rm D}=\mu_{\rm D}-\mu_{\rm p}-\mu_{\rm n}$, where 
$\mu_{\rm D}=0.857$, $\mu_{\rm p}=1+\kappa_{\rm p}=2.793$ and $\mu_{\rm n}=\kappa_{\rm n}=-1.913$ are the magnetic dipole moments of 
the physical deuteron, proton and neutron, respectively, and the 
electric quadrupole moment $(Q_{\rm D})_{\rm exp}=0.286\,{\rm fm}^2$ [5]. The magnetic dipole moment of the deuteron is measured in nuclear magnetons $\mu_{\rm N}\,=\,e/(2 M_{\rm N})$, where $e$ and $M_{\rm N}$ are the electric charge of the proton and the mass of the proton and neutron, then $\kappa_{\rm p}$ and $\kappa_{\rm n}$ are the anomalous magnetic dipole moments of the proton and neutron. In the case of the neutron the total magnetic dipole moment coincides with the anomalous one. Below, to simplify matter, we neglect the proton--neutron mass difference and use equal masses of the proton and neutron, i.e. $M_{\rm p}\,=\,M_{\rm n}\,=\,M_{\rm N}\,=\,938\,{\rm MeV}$. This should correspond to the chiral limit approximation with zero masses of current $u$-- and $d$--quarks, i.e. $m_{0u}\,=\,m_{0d}\,=\,0$. 

In this connection we would like to refer to the paper written by Sakita and Goebel [10], where, for the aim of the investigation of the low--energy limit of the photodisintegration, the deuteron has been described in terms of a local field operator.

The Lagrangian of the unphysical deuteron field $D^{(0)}_\mu 
(x)$, which interacts strongly with the proton $p(x)$ and neutron $n(x)$ 
fields, reads
\begin{eqnarray}\label{label1}
{\cal L}_{\rm st}(x) & = & -\frac{1}{2} D^{\dagger(0)}_{\mu\nu}(x) 
D^{(0)\mu\nu}(x) + M^2_0 D^{\dagger(0)}_\mu(x) D^{(0)\mu}(x) - 
\nonumber\\
&& - ig_{\rm V} \left[ \bar{p}(x) \gamma^\mu n^c(x)-\bar{n}(x) 
\gamma^\mu p^c(x) \right] D^{(0)}_\mu (x) - \nonumber\\
&& - ig_{\rm V} [ \bar{p^c}(x) \gamma^\mu n(x) - \bar{n^c}(x) \gamma^\mu 
p(x)] D^{\dagger(0)}_\mu (x) + \\
&& + \frac{g_{\rm T}}{M_0} [ \bar{p}(x) \sigma^{\mu\nu}  
n^c(x) - \bar{n}(x) \sigma^{\mu\nu} p^c(x)] 
D^{(0)}_{\mu\nu}(x) +\nonumber\\
&& +\frac{g_{\rm T}}{M_0} [ \bar{p^c}(x) \sigma^{\mu\nu} n(x)- 
\bar{n^c}(x) \sigma^{\mu\nu} p(x) ] D^{\dagger(0)}_{\mu\nu}(x) + 
\nonumber\\
&& + \bar{p}(x) (i\gamma^\mu \partial_\mu - M_{\rm N}) p(x) + \bar{n}(x) 
(i\gamma^\mu \partial_\mu - M_{\rm N}) n(x).\nonumber
\end{eqnarray}
Here $D^{(0)}_{\mu\nu}(x)=\partial_\mu D^{(0)}_\nu(x)-\partial_\nu D^{(0)}_\mu(x)$, $M_0=2M_{\rm N}$ is the mass of the unphysical deuteron, then $\psi^c(x) \,=\,C\,\bar{\psi}^T(x)$ and $\bar{\psi^c}(x) \,=\psi^T(x) \,C$, and  $\,\sigma^{\,\mu\,\nu}\,=\,\frac{1}{2}\,[\,\gamma^{\,\mu},\,\gamma^{\,\nu}]\,$. The operator $C$ denotes charge conjugation, $T$ is a transposition, $g_{\rm V}$ and $g_{\rm T}$ are the phenomenological constants which will be fixed below. We assume that the coupling constants $g_{\rm V}$ and $g_{\rm T}$ are caused by one--meson exchanges. The $\pi$--meson exchange should give the main contribution.

The guiding principle for the construction of the Lagrangian (\ref{label1}) has been only the fact that the deuteron is a bound state of the proton and neutron with spin one [2]. 

In order to compute the electric quadrupole and anomalous magnetic dipole moments of the physical deuteron we have to include the interaction the unphysical deuteron field $D^{(0)}_{\mu}(x)$ with the electromagnetic field. Having performed this inclusion by a minimal coupling we obtain 
the Lagrangian
\begin{eqnarray}\label{label2}
{\cal L}_{\rm tot}\,(x) = {\cal L}_{\rm st}\,(x) + {\cal L}_{\rm 
el}\,(x),
\end{eqnarray}
where
\begin{eqnarray}\label{label3}
{\cal L}_{\rm el}(x) & = & -ie\,D^{\dagger(0)}_{\mu\nu}(x) A^\mu (x) 
D^{(0)\nu}(x) + ie\,D^{(0)}_{\mu\nu}(x) A^\mu (x) D^{\dagger(0)\nu}(x) + 
\nonumber\\
&& + ie\,\frac{2g_{\rm T}}{M_0} [ \bar{p}(x) \sigma^{\mu\nu}  
n^c(x) - \bar{n}(x) \sigma^{\mu\nu} p^c(x) ] A_\mu(x) 
D^{(0)}_{\nu}(x) -\nonumber\\
&& - ie\frac{2g_{\rm T}}{M_0} [ \bar{p^c}(x) \sigma^{\mu\nu} n(x)- 
\bar{n^c}(x) \sigma^{\mu\nu} p(x) ] A_\mu(x) D^{\dagger(0)}_{\nu}(x) 
- \\
&&-i\,e\,\frac{\lambda}{M^2_{0}}\,D^{\dagger (0)}_{\mu\nu}(x)\,D^{(0)\nu \alpha}(x)\,{F_{\alpha}}^{\mu}(x)\,- e\,\bar{p}(x) \gamma^\mu p(x) A_\mu(x)\,+\nonumber\\
&& + ie\,\frac{\kappa_{\rm p}}{4M_{\rm N}}\,\bar{p}(x)\,\sigma^{\mu\nu}p(x)\,F_{\mu\nu}(x)\,+ \,ie\,\frac{\kappa_{\rm n}}{4M_{\rm N}} \bar{n}(x)\,\sigma^{\mu\nu}n(x)\,F_{\mu\nu}(x)+O(e^2).\nonumber
\end{eqnarray}
Here ${F_{\alpha}}^{\mu}(x) = \partial_{\alpha}A^{\mu}(x) - \partial^{\mu} A_{\alpha}(x)$ and $A_\mu(x)$ are the electromagnetic field strength tensor and the electromagnetic potential, respectively. Also we have added the Aronson interaction [11] describing the electric quadrupole and anomalous magnetic moments of the unphysical deuteron: $Q_{\rm D^{(0)}} = 2 \lambda/M^2_{0}$ and $\kappa_{\rm D^{(0)}} = \lambda$ [11]. 

Now let us give some arguments for the justification of the application of the Aronson interaction. Recall, that the electric quadrupole and the anomalous magnetic dipole moments characterise the inhomogeneity of the structure of the unphysical deuteron represented in terms of the field $D^{(0)}_{\mu}(x)$. The availability of such an inhomogeneity is not obvious beforehand and does not follow from the Lagrangian (\ref{label1}) via a minimal coupling inclusion of the interaction of the unphysical deuteron with an external electromagnetic field. Therefore, the inclusion of nonzero $Q_{\rm D^{(0)}}$ and $\kappa_{\rm D^{(0)}}$ is an additional assumption concerning the internal structure of the unphysical deuteron field.
Since, the minimal coupling leads to the appearance of electromagnetic interactions linear in $D^{(0)}_{\mu\nu}(x)$, the required interaction of the unphysical deuteron field with an external electromagnetic field, caused by electric quadrupole and anomalous magnetic dipole moments, should be of order $O(D^{(0)}_{\mu\nu}(x) D^{(0)}_{\alpha\beta}(x))$. This does not contradict to the main principles of Classical Electrodynamics [12] and gives the required interaction in the form, being irreducible with respect to those interactions induced by a minimal coupling. The Aronson effective Lagrangian provides a simplest form of an interaction of order $O(D^{(0)}_{\mu\nu}(x)D^{(0)}_{\alpha\beta}(x))$ [11]. 

The inclusion of the Aronson interaction can be also justified by the assumption that the electromagnetic interactions induced by electric quadrupole and anomalous magnetic dipole moments of the unphysical deuteron field with an external electromagnetic field become available only in next--to--leading order in long--wavelength expansion, i.e. in powers of gradients of the deuteron field $D^{(0)}_{\mu\nu}(x)$. In leading order in the long--wavelength expansion, linear in  $D^{(0)}_{\mu\nu}(x)$, the interaction of the unphysical deuteron with an external electromagnetic field, caused by electric quadrupole and anomalous magnetic dipole moments, vanishes.

We have taken into account the anomalous magnetic dipole moments of the proton and neutron that play an important role for the radiative neutron--proton capture n + p $\to$ D + $\gamma$. The inclusion of $\lambda$ and the anomalous magnetic dipole moments of the proton and neutron distinguishes the present model from that given in [1].

Some of the results and assertions adduced in this section repeat those given in [1]. This concerns mainly the computation of the binding energy $\varepsilon_{\rm D}$ and the discussion of the validity of the approximation having been applied to the computation of one--nucleon loop diagrams. This repetition is caused by the consideration of the completeness of the exposition, i.e. in order not to relegate a reader to [1] for consultation, and the change of the magnitudes of phenomenological parameters of the model. The former is due to different procedures of the removal of ambiguities of one--nucleon loop diagrams, induced by shifts of virtual nucleon momenta, we use here and we have used in Ref.~[1]. In the present version all shift ambiguities are fully fixed by requirements of gauge invariance.

Following [1] we compute the effective Lagrangian of the physical deuteron field in the one--nucleon loop approximation (see also [6--9]). The nucleon 
diagrams describing the contributions to the kinetic term of the 
deuteron field and leading to the appearance of the non--zero value of 
the binding energy are depicted in Fig.~1. By calculating these diagrams in leading order in the long--wavelength expansion we get

\parbox{11cm}{\begin{eqnarray*}
\delta {\cal L}_{\rm eff}(x) &=& -\frac{1}{2} \Bigg[\frac{g^2_{\rm V} - 6 g_{\rm V}g_{\rm T} + 3 
g^2_{\rm T}}{3\pi^2} J_2(M_{\rm N}) \Bigg]\, D^{\dagger(0)}_{\mu\nu}(x)\,D^{(0)\mu\nu}(x)  \,-\\
&&-\,\frac{2}{3}\,\frac{g^2_{\rm V}}{\pi^2}\,\Big[J_1(M_{\rm N})\,+\,M^2_{\rm N}\,J_2(M_{\rm N})\Big]\, D^{\dagger(0)}_{\mu}(x)\,D^{(0)\mu}(x),  
\end{eqnarray*}} \hfill
\parbox{1cm}{\begin{eqnarray}\label{label4}
\end{eqnarray}}

\noindent where $J_1(M_{\rm N})$ and $J_2(M_{\rm N})$ are the following divergent integrals

\parbox{11cm}{\begin{eqnarray*}
J_1(M_{\rm N}) &=&\int\frac{d^4k}{\pi^2i}\frac{1}{M_{\rm N}^2 - k^2}= 4\int^{\Lambda}_{0}\frac{d|\vec{k}|{\vec{k}}^{\,2}}{(M^2_{\rm N}\,+\,{\vec{k}}^{\,2})^{\,1/2}}\,,\\
J_2(M_{\rm N}) &=&\int\frac{d^4k}{\pi^2i}\frac{1}{(M_{\rm N}^2 - k^2)^{\,2}}= 2\int^{\Lambda}_{0}\frac{d|\vec{k}|{\vec{k}}^{\,2}}{(M^2_{\rm N}\,+\,{\vec{k}}^{\,2})^{\,3/2}}\,.
\end{eqnarray*}} \hfill
\parbox{1cm}{\begin{eqnarray}\label{label5}
\end{eqnarray}}

\noindent The ultra--violet cut--off $\Lambda$ restricts the 3--momenta of fluctuations of virtual nucleons taking part in the 
formation of the physical deuteron field. One should expect that 
$\Lambda$ is connected with the region of localization of a wave 
packet procreated by fluctuations of virtual nucleons, i.e. $\Delta 
r\,\cdot\,\Lambda \approx 1$. We shall specify the value of $\,\Lambda\,$ below.

Now we discuss the problem of the applicability of the 
long--wavelength expansion for the calculation of the 
one--nucleon loop diagrams in Fig.~1. It is well--known that a 
two--body $S$--wave bound state being denoted as $D$ with a 
reduced mass $M_{\rm N}/2$  and a binding energy $\varepsilon_{\rm D}$ is localized in 
the region restricted by $r_{\rm D} =1/\sqrt{\varepsilon_{\rm D}M_{\rm N}}$ [4]. This 
quantity can be considered as the effective radius of the bound state 
[5]. In the case of the physical deuteron we have $r_{\rm D} = 
4.319\,{\rm fm}$ [5]. This value exceeds three times the effective radius of 
nuclear forces $r_{\rm N}=1/M_\pi=1.462\,{\rm fm}$, where 
$M_\pi=134.976\,{\rm MeV}$ is the pion mass [13]. Hence the physical deuteron looks like a rather "extended" bound state. It should be obvious that such an "extended" bound state can be formed at the expense of the main contributions of long--wavelength fluctuations of the bound proton and neutron. It implies that we can expect a cut--off $\Lambda$ satisfying the inequality $\Lambda\ll M_{\rm N}$. It should be obvious that $\Lambda$ is to be identified with $\Lambda_{\rm D}\,=\,1/r_{\rm D}\,=\,45.688\;{\rm MeV}\,$, i.e. $\Lambda\,=\,\Lambda_{\rm D}\,=\,45.688\;{\rm MeV}\,$.

The infinitesimality of the derivatives $\partial_\nu D^{(0)}_\mu(x)$ necessary for the validity of the long--wavelength expansion
can be justified as follows. We assume that the deuteron field 
$D^{(0)}_\mu(x)$ is a bound state of a proton and a neutron at zero 
binding energy. As has been mentioned above, this means that the field 
$D^{(0)}_\mu(x)$ is localized in the region whose upper boundary 
$r_{\rm D^{(0)}}$ goes to infinity. Obviously it results in 
a smooth variation of the field $D^{(0)}_\mu(x)$ at the scale of the 
effective radius of the physical deuteron. Thus we have adduced some 
arguments on behalf of the validity of the long--wavelength expansion 
which has been applied for the calculation of the effective Lagrangian 
(\ref{label4}).

Now we proceed to the computation of the binding energy of the physical deuteron. First, following [1], we introduce the field of the physical deuteron
\begin{eqnarray}\label{label6}
D_\mu(x)=Z^{1/2}_{\rm D} D^{(0)}_\mu(x),
\end{eqnarray}
where 
\begin{eqnarray}\label{label7}
Z_{\rm D} = 1 + \frac{g^2_{\rm V}-6g_{\rm V}g_{\rm T}+3 
g^2_{\rm T}}{3\,\pi^2}\, J_2(M_{\rm N})
\end{eqnarray}
is the wave--function normalization constant. After the 
renormalization (7) we get the effective Lagrangian of the free physical 
deuteron field $D_\mu(x)$ [1]
\begin{eqnarray}\label{label8}
{\cal L}^{(0)}_{\rm eff}(x) = -\frac{1}{2}\, D^{\dagger}_{\mu\nu}(x) 
D^{\mu\nu}(x) + M^2_{\rm D}\, D^{\dagger}_\mu(x) D^\mu(x),
\end{eqnarray}
where $M_{\rm D} = M_0-\varepsilon_{\rm D}$ is the mass of the physical deuteron 
and $\varepsilon_{\rm D}$ is the binding energy. In the one--nucleon loop approximation and in leading order of the long--wavelength expansion the binding energy $\varepsilon_{\rm D}$ is determined by the expression 
\begin{eqnarray}\label{label9}
\varepsilon_{\rm D} = \,\frac{g^2_{\rm V}}{6\,\pi^2}\,\frac{1}{M_{\rm N}}\,J_1(M_{\rm N})\,+
\,\frac{g^2_{\rm V}-4g_{\rm V}g_{\rm T}+2g^2_{\rm T}}{2\pi^2}\,M_{\rm N}\, J_2(M_{\rm N}).
\end{eqnarray}
For the derivation of Eq.~(\ref{label9}) we have used the 
inequality 
\begin{eqnarray}\label{label10}
 \frac{g^2_{\rm V}-6g_{\rm V}g_{\rm T}+3g^2_{\rm T}}{3\,\pi^2}\, 
 J_2(M_{\rm N})\ll 1
\end{eqnarray}
being consistent with the inequality $\Lambda_{\rm D}\ll M_{\rm N}$  discussed above. For $\Lambda_{\rm D}\ll M_{\rm N}$  Eq.~(\ref{label9}) reads 
\begin{eqnarray}\label{label11}
\varepsilon_{\rm D} \,=\,\Lambda^3_{\rm D}\,\frac{5}{9}\frac{g^2_{\rm V}}
{\pi^2}\,\frac{1}{M^2_{\rm N}}\Bigg[1\,-\,\frac{12}{5}\,\Bigg(\frac {g_{\rm T}}{g_{\rm V}}\Bigg)\, +\,\frac{6}{5}\,\Bigg(\frac {g_{\rm T}}{g_{\rm V}}\Bigg)^2\Bigg]\,.
\end{eqnarray}
It is seen that the binding energy of the physical deuteron is 
expressed in terms of three phenomenological parameters of our model: 
$\Lambda_{\rm D},\;g_{\rm V}$ and $g_{\rm T}$. Then we have to fix the coupling constants $g_{\rm V}$ and $g_{\rm T}$ to obtain the theoretical value of the binding energy.

Eqs.~(\ref{label9}) and (\ref{label11}) have been calculated in [1]. We need to reproduce them here for two reasons. First, we give below the new relation between the coupling constant $g_{\rm V}$ and the electric quadrupole moment $Q_{\rm D}$. Second, in the Conclusion we shall discuss Eq.~(\ref{label11}) setting $g_{\rm T}=0$. For these reasons it is convenient to have Eqs.~(\ref{label9}) and (\ref{label11}) at hand and not to relegate a reader to Ref.~[1]. 

In order to fix the coupling constants $g_{\rm V}$ and $g_{\rm T}$ we suggest to calculate the anomalous magnetic dipole $\kappa_{\rm D}$  and electric quadrupole  $Q_{\rm D}$ moments in terms of $g_{\rm V}$ and $g_{\rm T}$. We admit that $\kappa_{\rm D}$  and  $Q_{\rm D}$ appear in the one--nucleon loop approximation, and only at the expense of interactions taken into account in the Lagrangian (\ref{label2}) [1]. The complete set of one--nucleon loop diagrams is depicted in Fig.~2. The non--trivial contributions come from the diagrams in Figs.~2c,d. The detailed computation of the diagrams depicted in Figs.~2c,d is given in Sects.~2 and 3, respectively. Here we only adduce the results.

The effective Lagrangian, defined by the one--nucleon loop diagrams in Fig.~2 and describing the electromagnetic interactions of the physical deuteron field via the anomalous magnetic and electric quadrupole moments, is given by
\begin{eqnarray}\label{label12}
{\cal L}^{\rm el}_{\rm eff}(x)&=&ie\frac{g^2_{\rm 
V}}{2\pi^2}\,D^{\dagger}_\mu(x) D_\nu(x) F^{\mu\nu}(x) +\\ &+&ie\Bigg\{\frac{2\,g^2_{\rm T}}{3\pi^2}\Bigg[1-\frac{9}{8}(\kappa_{\rm p}-\kappa_{\rm n})\Bigg(1-\frac{8}{9}\frac{g_{\rm V}}{g_{\rm T}}\Bigg)\Bigg]-\lambda\Bigg\}\frac{1} 
{M^2_{\rm D}}\,D^{\dagger}_{\mu\nu}(x) D^{\nu\alpha}(x) 
{F_{\alpha}}^{\mu}(x).\nonumber
\end{eqnarray}
The Lagrangian (\ref{label12}) contains only  physical deuteron fields. The first term in 
${\cal L}^{\rm el}_{\rm eff}(x)$ is the Corben--Schwinger 
interaction [14], while the second term presents the interaction which has first been introduced by Aronson [11]. These interactions describe the anomalous magnetic dipole and electric quadrupole moments of the charged vector field. It should be emphasized that we have neglected the divergent contributions which are small in comparison with the convergent ones due to the restriction (\ref{label10}).

Note that the effective Lagrangian (\ref{label12}) differs from that obtained in Ref.~[1] by the additional contributions caused by anomalous magnatic moments of the proton and neutron, and the  phenomenological coupling $\lambda$ induced by the Aronson interaction. If one would set $\kappa_{\rm p} = \kappa_{\rm n} = \lambda = 0$, one would find a discrepancy in the main terms.The former is due to different procedures of the removal of ambiguities of one--nucleon loop diagrams, induced by shifts of virtual nucleon momenta, we use here and we have used in Ref.~[1]. In the present version all shift ambiguities are fully fixed by requirements of gauge invariance.

The anomalous magnetic dipole moment $\kappa_{\rm D}$ measured in units of the nuclear 
magneton $\mu_{\rm N} = e/(2 M_{\rm N})$ and the electric quadrupole moment $Q_{\rm D}$ are given by [10,13]
\begin{eqnarray}\label{label13}
\kappa_{\rm D}\,=\,-\,\frac{g^2_{\rm V}}{4\,\pi^2}\,-\,\frac{1}{2}\,\Bigg\{\frac{2\,g^2_{\rm T}}{3\pi^2}\Bigg[1-\frac{9}{8}(\kappa_{\rm p}-\kappa_{\rm n})\Bigg(1-\frac{8}{9}\frac{g_{\rm V}}{g_{\rm T}}\Bigg)\Bigg]\,-\,\lambda\Bigg\}\,,
\end{eqnarray}
\begin{eqnarray}\label{label14}
Q_{\rm D}\,=\,\Bigg\{\frac{g^2_{\rm V}}{\pi^2}\,-\,2\,\Bigg\{\frac{2\,g^2_{\rm T}}{3\pi^2}\Bigg[1-\frac{9}{8}(\kappa_{\rm p}-\kappa_{\rm n})\Bigg(1-\frac{8}{9}\frac{g_{\rm V}}{g_{\rm T}}\Bigg)\Bigg]\,-\,\lambda\Bigg\}\Bigg\}\frac{1}{M^2_{\rm D}}\,.
\end{eqnarray}
The experimental value $(\kappa_{\rm D})_{\rm exp}=-0.023$ of the anomalous magnetic dipole moment of the deuteron  is small compared with the magnetic dipole moment of the deuteron $\mu_{\rm D}\,=\,0.857$. Therefore, we suggest to set $\kappa_{\rm D}=0$ giving
\begin{eqnarray}\label{label15}
\frac{g^2_{\rm V}}{2\pi^2}\,=\,\lambda\,-\,\frac{2\,g^2_{\rm T}}{3\pi^2}\Bigg[1-\frac{9}{8}(\kappa_{\rm p}-\kappa_{\rm n})\Bigg(1-\frac{8}{9}\frac{g_{\rm V}}{g_{\rm T}}\Bigg)\Bigg]\,.
\end{eqnarray}
In this case the observed magnitude $(\kappa_{\rm D})_{\rm exp}=-0.023$ can be the matter of next--to--leading order corrections to the accepted approximation, for example, chiral corrections, etc.

By applying Eq.~(\ref{label15}) we can express the electric quadrupole moment 
$Q_{\rm D}$ in terms of the coupling constant $g_{\rm V}$ only, i.e.
\begin{eqnarray}\label{label16}
Q_{\rm D} = \frac{2g^2_{\rm V} }{\pi^2} \frac{1}{M^2_{\rm D}}\,.
\end{eqnarray}
The relation between $g_{\rm V}$ and $Q_{\rm D}$, represented by Eq.~(\ref{label16}), differs by a factor 2/3 from that obtained in Ref.~[1]. 

Such a discrepancy has the following explanation. The relation Eq.~(\ref{label16}) is caused by effective couplings of the deuteron to electromagnetic field, i.e. the Corben--Schwinger and Aronson interactions. These interactions are defined by one--nucleon loop diagrams the computation of which depends substantially on the shift of virual nucleon momenta (see Sects.~3 and 4). The mathematical procedure of the removal of such ambiguities which has been used in Ref.[1], unfortunately, has left room for residual ambiguities. In the present version ambiguities induced by shifts of virtual nucleon momenta are fully fixed by requirements of gauge invariance under gauge transformations of the electromagnetic and the deuteron fields. The former has led to the factor 2/3 in the relation between $Q_{\rm D}$ and $g^2_{\rm V}$.

By using the experimental value $(Q_{\rm D})_{\rm exp}=0.286\,{\rm fm^2}$ and 
$M_{\rm D}\simeq 1876\,{\rm MeV}$ we can estimate the value of $g_{\rm V}$. This gives: $g_{\rm V} = \pm 11.3$. Without loss of generality we can use the positive sign, i.e.
\begin{eqnarray}\label{label17}
g_{\rm V} = 11.3.
\end{eqnarray}
The coupling constant $g_{\rm V}$ satisfies the relation $g_{\rm V}\simeq g_{\pi\rm NN}$ , where $g_{\pi\rm NN}\,=\,13.4\pm\,0.1$ [4] is the coupling constant of the ${\pi\rm NN}$--interactions. Thus the magnitude of $g_{\rm V}$ corroborates our assumption that the phenomenological interactions of the deuteron and the proton and neuteron, given in the Lagrangian (\ref{label1}), are caused by one--meson exchange and the $\pi$--meson exchange gives the main contribution. This admission can be justified by comparing the effective radii of one--meson exchanges. The radii of pion, $\sigma\,(660)\,$--meson [6,9] and $\rho\,(770)\,$--meson exchanges are defined in terms of their masses, i.e. $r_{\pi}\,=\,1/M_{\pi}\,=\,1.462\,{\rm fm}$ at $M_{\pi}=134.976\,{\rm MeV}\,$ [13], $r_{\sigma} \,=\, 1/M_{\sigma} \,=\, 0.30\,{\rm fm}$ at $M_{\sigma} \,=\, 660\;{\rm MeV}\,$ [5,9] and $r_{\rho} \,=\, 1/M_{\,\rho} \,=\, 0.27\;{\rm fm}$ at $M_{\,\rho} \,=\, 770\,{\rm MeV}\,$, respectively. The radii of the $\sigma\,(660)\,$ and $\rho\,(770)\,$--meson exchanges are much smaller than the radius of the $\pi$--meson exchange. Therefore, these interactions of the proton and neutron with $\sigma\,(660)\,$ and $\rho\,(770)\,$ mesons should contribute perturbatively to the deuteron problem. 

We have put $\kappa_{\rm D}\,=\,0$ as the experimental value $(\kappa_{\rm D})_{\exp}\,=\,-\,0.023$ is small enough in comparison with the magnetic dipole moment of the deuteron $\mu_{\rm D}\,=\,0.857$. We assume that the nonzero value of $\kappa_{\rm D}$ can be obtained perturbatively by taking into account nonzero values of current quark masses, for example, within Chiral perturbation theory at the quark level (CHPT)$_q$ [9], based on the extended Nambu--Jona--Lasinio model with linear realization of chiral $U(3)\times U(3)$ symmetry, and interactions of the proton and neutron with $\sigma\,(660)\,$ and $\rho\,(770)\,$--mesons within the one--meson exchange approximation.

From Eqs.~(\ref{label11}) and (\ref{label16}) we can express 
the binding energy $\varepsilon_{\rm D}$ in terms of the electric quadrupole 
moment $Q_{\rm D}$ and the ratio $g_{\rm V}/g_{\rm T}$ 
\begin{eqnarray}\label{label18}
\varepsilon_{\rm D} = \frac{10}{9}\, Q_{\rm D} \,\Lambda^3_{\rm D}\, \Bigg[1\,-\,\frac{12}{5}\,\Bigg(\frac{g_{\rm T}}{g_{\rm V}}\Bigg)\, +\,\frac{6}{5}\,\Bigg(\frac{g_{\rm T}}{g_{\rm V}}\Bigg)^{\,2}\Bigg]\,.
\end{eqnarray}
This formula agrees with the statement that the main contribution to 
the binding energy of the deuteron comes from the tensor forces producing the non--zero value of the quadrupole moment $Q_{\rm D}$ [3]. For the computation of the magnitude of the binding energy we have to fix the ratio $g_{\rm T}/g_{\rm V}$ that is still free. Putting $g_{\rm T}/g_{\rm V}\,=\,-\,\sqrt{3/8}$ [1]  we obtain the best fit of the binding energy
\begin{eqnarray}\label{label19}
\varepsilon_{\rm D}\,=\, 2.273\,{\rm MeV}\,.
\end{eqnarray}
The accuracy of the fit makes up 2$\%$. Thus, we have fixed all phenomenological parameters and described all quantities, characterizing the physical deuteron.

The total effective Lagrangian of the physical deuteron describing strong and electromagnetic interactions of the deuteron reads 
\begin{eqnarray}\label{label20}
{\cal L}_{\rm tot}(x) &=&-\frac{1}{2} D^{\dagger}_{\mu\nu}(x) 
D^{\mu\nu}(x) + M^2_D D^{\dagger}_\mu(x) D^{\mu}(x) -ieD^{\dagger}_{\mu\nu}(x) A^\mu (x) D^{\nu}(x) + \nonumber\\
&&+ieD_{\mu\nu}(x) A^\mu (x) D^{\dagger\nu}(x) +ie\Bigg[\frac{g^2_{\rm V}}{2\pi^2}\Bigg] D^{\dagger}_\mu(x) 
D_\nu(x) F^{\mu\nu}(x) -\nonumber\\
&&-ie\Bigg[\frac{g^2_{\rm V}}{2\pi^2}\Bigg] \,\frac{1} 
{M^2_{\rm D}}\,D^{\dagger}_{\mu\nu}(x) D^{\nu\alpha}(x) 
{F_{\alpha}}^{\mu}(x) -\nonumber\\
&&-ig_{\rm V} [ \bar{p}(x) \gamma^\mu n^c(x)-\bar{n}(x) 
\gamma^\mu p^c(x)] D_\mu (x) - \nonumber\\
&&-ig_{\rm V} [ \bar{p^c}(x) \gamma^\mu n(x) - \bar{n^c}(x) \gamma^\mu 
p(x) ] D^{\dagger}_\mu (x) + \\
&&+\frac{g_{\rm T}}{M_{\rm D}} [ \bar{p}(x) \sigma^{\mu\nu}  
n^c(x)  - \bar{n}(x) \sigma^{\mu\nu} p^c(x)  ] D_{\mu\nu}(x) +\nonumber\\ 
&&+\frac{g_{\rm T}}{M_{\rm D}} [ \bar{p^c}(x)  \sigma^{\mu\nu} n(x)- 
\bar{n^c}(x)  \sigma^{\mu\nu} p(x) ] D^{\dagger}_{\mu\nu}(x) + 
\nonumber\\
&&+ ie\frac{2g_{\rm T}}{M_{\rm D}} [ \bar{p}(x) \sigma^{\mu\nu}  
n^c(x)  - \bar{n}(x) \sigma^{\mu\nu}  p^c(x)  ] A_\mu(x) 
D_{\nu}(x) - \nonumber\\
&&- ie\frac{2g_{\rm T}}{M_{\rm D}} [ \bar{p^c}(x)  \sigma^{\mu\nu} n(x)- 
\bar{n^c}(x)  \sigma^{\mu\nu} p(x) ] A_\mu(x) D^{\dagger}_{\nu}(x) + O(e^2) + ... .\nonumber
\end{eqnarray}
The ellipses stand for interactions of the proton and neutron with other fields such as photon, pions, etc. The effective Lagrangian (\ref{label20}) differs from that obtained in Ref.~[1] by the coupling constants of the effective interactions of the physical deuteron with the electromagnetic field describing the anomalous magnetic dipole $\kappa_{\rm D}$ and electric quadrupole moment $Q_{\rm D}$.

We have to underline that the suggested field theory model of the deuteron is applicable only at the low--energy limit. Thereby all interactions of the deuteron with other hadrons should run via one--nucleon  loop exchange. This is due to the one--nucleon  loop origin of the deuteron in our model. This assertion is very similar to the approximation accepted within the NJL model, where all interactions of hadrons run via one--constituent quark--loop exchange [6--9]. However, if the one--constituent quark loop approximation in the extended NJL quark model [9] can be justfied in large N (number of quark colours) expansion, where the perturbation theory is developed in powers of $1/{\rm N}$. In the case of the  relativistic field theory model of the deuteron for the justification of the one--nucleon loop approximation  we do not have any perturbative parameter. Therefore, our approximation cannot be justified like that in the extended NJL quark model and should be considered as an effective one. For the confirmation of the validity of this effective approximation we can only refer to Ref.[1], where the two--nucleon loop contributions to the binding energy have been calculated and found much less than the one--nucleon loop contributions.

We deem that most likely the deuteron cannot be inserted in an intermediate state of any  process of low--energy interactions. This is connected with a very sensitive structure of the deuteron as an "extended" bound state with a small binding energy. The representation of such a state in terms of any local quantum field is rather limited. The latter entails an undetermined character of the description of the deuteron in intermediate states in terms of Green functions of these local fields.

Now we can proceed to the computation of the effective interactions describing the anomalous magnetic and electric quadrupole moments of the deuteron. The complete set of one--nucleon loop diagrams describing effective interaction of the deuteron with the electromagnetic field is depicted in Fig.~2. In leading order in the long--wavelength expansion the diagrams in Figs.~2a,b do not contribute to the anomalous magnetic and electric quadrupole moments. The contributions of these diagrams are divergent and can be removed by the renormalization of the electric charge of the deuteron. The nontrivial contributions to the effective anomalous magnetic and electric quadrupole moments are defined only by the diagrams in Figs.~2c,d.

\section{The effective Corben--Schwinger interaction}
\setcounter{equation}{0}

\hspace{0.2in} In this Section we give the detailed derivation of the effective Corben--Schwinger Lagrangian defined by the one--nucleon loop diagram in Fig.~2c and describing the effective electromagnetic interactions of the deuteron. The effective Lagrangian of the diagram in Fig.~2c is defined
\begin{eqnarray}\label{label21}
\int\,d^4x\,{\cal L}_{\rm Fig.2c}(x)&=&\int\,d^4x\,\int\,\frac{d^4x_1\,d^4k_1}{(2\pi)^4}\,\frac{d^4x_2\,d^4k_2}{(2\pi)^4}\,D_{\beta}(x)\,D^{\dagger}_{\alpha}(x_1)\,A_{\mu}(x_2)\,\times\nonumber\\
&&\times \, e^{-\,i\,k_1\cdot x_1}\,e^{-\,i\,k_2\cdot x_2}\,e^{\,i\,(k_1\,+\,k_2)\cdot x}\,\frac{e\,g^2_{\rm V}}{4\,\pi^2}\,{\cal J}^{\beta\alpha\mu}(k_1, k_2; Q)\,,
\end{eqnarray}
where
\begin{eqnarray}\label{label22}
&&{\cal J}^{\beta\alpha\mu}(k_1, k_2; Q)\,=\\
&&=\,\int\frac{d^4k}{\pi^2i}\,{\rm tr}\Bigg\{\gamma^{\beta}\,\frac{1}{M_{\rm N}-\hat{k}-\hat{Q}}\,\gamma^{\alpha}\,\frac{1}{M_{\rm N}-\hat{k}-\hat{Q}-\hat{k}_1}\,\gamma^{\mu}\,\frac{1}{M_{\rm N}-\hat{k}-\hat{Q}-\hat{k}_1-\hat{k}_2}\Bigg\}\,.\nonumber
\end{eqnarray}
The 4--vector $Q\,=\,a\,k_1\,+\,b\,k_2$, where $a$ and $b$ are arbitrary constants, displays the dependence of the $k$ integral in (\ref{label21}) on the shift of the virtual momentum. This ambiguity of the computation of the integral over $k$, which has been found by Gertsein and Jackiw [15], is used to remove undesirable contributions and make the effective Lagrangian gauge invariant. We use the fields of the physical deuteron. This is because the renormalization (\ref{label6}) introduces divergent terms that are small compared with the convergent ones we are following for the computation of the Corben--Schwinger Lagrangian.

In order to display the Gertsein--Jackiw ambiguity we follow the Gertsein--Jackiw method and compute the difference
\begin{eqnarray}\label{label23}
\delta\,{\cal J}^{\beta\alpha\mu}(k_1, k_2; Q)\,=\,{\cal J}^{\beta\alpha\mu}(k_1, k_2; Q)\,-\,{\cal J}^{\beta\alpha\mu}(k_1, k_2; 0)\,.
\end{eqnarray}
In accordance with the Gertsein--Jackiw method the difference (\ref{label23}) can be represented by the integral
\begin{eqnarray}\label{label24}
&&\delta\,{\cal J}^{\beta\alpha\mu}(k_1, k_2; Q)\,=\,\int^{1}_{0}\,dx\,\frac{d}{dx}{\cal J}^{\beta\alpha\mu}(k_1, k_2; x Q)\,=\nonumber\\
&&= \int^{1}_{0}\,dx\,\int\frac{d^4k}{\pi^2i}\,Q^{\lambda}\,\frac{\partial}{\partial\,k^{\lambda}}\,{\rm tr}\Bigg\{\gamma^{\beta}\,\frac{1}{M_{\rm N}-\hat{k}-x\hat{Q}}\,\gamma^{\alpha}\,\frac{1}{M_{\rm N}-\hat{k}-x\hat{Q}-\hat{k}_1}\,\gamma^{\mu}\,\nonumber\\
&&\times \,\frac{1}{M_{\rm N}-\hat{k}-x\hat{Q}-\hat{k}_1-\hat{k}_2}\Bigg\}\,.
\end{eqnarray}
This shows that the Gertsein--Jackiw ambiguity is just the surface term. Following Gertsein and Jackiw [15] and evaluating the integral over $k$ symmetrically, we obtain
\begin{eqnarray}\label{label25}
&&\delta\,{\cal J}^{\beta\alpha\mu}(k_1, k_2; Q)\,=\,-\,2\,\int^{1}_{0}\,dx\,\lim_{R\to \infty}\Bigg<\frac{Q\cdot R}{R^4}{\rm tr}\{\gamma^{\beta}\,(M_{\rm N}+\hat{R}+x\hat{Q})\,\gamma^{\alpha}\,\times \nonumber\\
&&\times (M_{\rm N}+\hat{R}+x\hat{Q}+\hat{k}_1)\,\gamma^{\mu}\,(M_{\rm N}+\hat{R}+x\hat{Q}+\hat{k}_1+\hat{k}_2)\}\Bigg>\,.
\end{eqnarray}
The brackets $<\ldots>$ denote the averaging over directions of the 4-vector $R$. Due to the limit $R\to \infty$ we can neglect all momenta with respect to $R$.
\begin{eqnarray}\label{label26}
&&\delta\,{\cal J}^{\beta\alpha\mu}(k_1, k_2; Q)\,=\,-\,2\,\lim_{R\to \infty}\Bigg<\frac{Q\cdot R}{R^4}{\rm tr}\{\gamma^{\beta}\hat{R}\,\gamma^{\alpha}\hat{R}\,\gamma^{\mu}\hat{R}\}\Bigg>\,.
\end{eqnarray}
Averaging over directions of the 4--vector $R$
\begin{eqnarray}\label{label27}
\lim_{R\to \infty}\frac{R^{\lambda}R^{\varphi}R^{\omega}R^{\rho}}{R^4}\,=\,\frac{1}{24}\,(g^{\lambda\varphi}\,g^{\omega\rho}\,+\,g^{\lambda\omega}\,g^{\varphi\rho}\,+\,g^{\lambda\rho}\,g^{\varphi\omega})\,,
\end{eqnarray}
we obtain
\begin{eqnarray}\label{label28}
&&\delta\,{\cal J}^{\beta\alpha\mu}(k_1, k_2; Q)\,=\,-\,\frac{1}{12}\,{\rm tr}(\gamma_{\lambda}\,\gamma^{\beta}\,\gamma^{\lambda}\,\gamma^{\alpha}\hat{Q}\,\gamma^{\mu}\,+\,\gamma^{\beta}\gamma_{\lambda}\gamma^{\alpha}\gamma^{\lambda}\hat{Q}\,\gamma^{\mu}\,+\\
&&+\,\gamma_{\beta}\hat{Q}\,\gamma^{\alpha}\gamma_{\lambda}\gamma^{\mu}\gamma^{\lambda})\,=\,\frac{2}{3}\,(Q^{\alpha}g^{\beta\mu}\,+\,Q^{\beta}g^{\mu\alpha}\,+\,Q^{\mu}g^{\alpha\beta}).\nonumber
\end{eqnarray}
Thus the surface ambiguity noticed by Gertsein and Jackiw contains only finite contributions.

Now we should proceed to the computation of ${\cal J}^{\beta\alpha\mu}(k_1, k_2; Q)$. In order to pick up the ambiguity connected with $Q$ one cannot apply the Feynman method of the computation of momentum integrals like (\ref{label22}). This method involves the merger of the factors in the denominator with the subsequent shift of virtual momentum. On this way one can lose the $Q$--depenedence that is due to the shift at the intermediate stage. Thereby  we have to compute the integral over $k$  without any intermediate shift. One can carry out this by applying a long--wavelength expansion and keeping to the leading terms.
\begin{eqnarray}\label{label29}
&&{\cal J}^{\beta\alpha\mu}(k_1, k_2; Q)\,=\nonumber\\
&&=\,\int\frac{d^4k}{\pi^2i}\,{\rm tr}\Bigg\{\gamma^{\beta}\,\frac{M_{\rm N}+\hat{k}+\hat{Q}}{M^2_{\rm N}-k^{\,2}}\,\Bigg[1\,+\,\frac{2\,k\cdot Q}{M^2_{\rm N}-k^{\,2}}\Bigg]\,
\gamma^{\alpha}\,\frac{M_{\rm N}+\hat{k}+\hat{Q}+\hat{k}_1}{M^2_{\rm N}-k^{\,2}}\,\times \nonumber\\
&&\times\,\Bigg[1\,+\,\frac{2\,k\cdot (Q+k_1)}{M^2_{\rm N}-k^{\,2}}\Bigg]\,\gamma^{\mu}\,\frac{M_{\rm N}+\hat{k}+\hat{Q}+\hat{k}_1+\hat{k}_2}{M^2_{\rm N}-k^{\,2}}\Bigg[1\,+\,\frac{2\,k\cdot (Q+k_1+k_2)}{M^2_{\rm N}-k^{\,2}}\Bigg]\Bigg\}\,=\nonumber\\
&&=\,\int\frac{d^4k}{\pi^2i}\,\frac{1}{(M^2_{\rm N}-k^{\,2})^{\,3}}{\rm tr}\,\{M^2_{\rm N}\gamma^{\beta}(\hat{k}+\hat{Q})\gamma^{\alpha}\gamma^{\mu}+M^2_{\rm N}\gamma^{\beta}\gamma^{\alpha}(\hat{k}+\hat{Q}+\hat{k}_1)\gamma^{\mu}+\nonumber\\
&&+M^2_{\rm N}\gamma^{\beta}\gamma^{\alpha}\gamma^{\mu}(\hat{k}+\hat{Q}+\hat{k}_1+\hat{k}_2)+\gamma^{\beta}(\hat{k}+\hat{Q})\gamma^{\alpha}(\hat{k}+\hat{Q}+\hat{k}_1)\gamma^{\mu}(\hat{k}+\hat{Q}+\hat{k}_1+\hat{k}_2)\}\nonumber\\
&&\Bigg[1\,+\,\frac{2\,k\cdot (3Q+2k_1+k_2)}{M^2_{\rm N}-k^{\,2}}\Bigg]\,=\\
&&=\,\frac{1}{2}\int\frac{d^4k}{\pi^2i}\Bigg[\frac{1}{(M^2_{\rm N}-k^2)^2}+\frac{M^2_{\rm N}}{(M^2_{\rm N}-k^2)^3}\Bigg]{\rm tr}\,\{\gamma^{\beta}\hat{Q}\gamma^{\alpha}\gamma^{\mu}+\gamma^{\beta}\gamma^{\alpha}(\hat{Q}+\hat{k}_1)\gamma^{\mu}+\nonumber\\
&&+\gamma^{\beta}\gamma^{\alpha}\gamma^{\mu}(\hat{Q}+\hat{k}_1+\hat{k}_2)\}\,+\nonumber\\
&&+\,2\,\int\frac{d^4k}{\pi^2i}\frac{k\cdot (3Q+2k_1+k_2)}{(M^2_{\rm N}-k^2)^4}{\rm tr}\,\{M^2_{\rm N}(\gamma^{\beta}\hat{k}\gamma^{\alpha}\gamma^{\mu}+\gamma^{\beta}\gamma^{\alpha}\hat{k}\gamma^{\mu}+\gamma^{\beta}\gamma^{\alpha}\gamma^{\mu}\hat{k})+\nonumber\\
&&+\gamma^{\beta}\hat{k}\gamma^{\alpha}\hat{k}\gamma^{\mu}\hat{k}\}\,=\,{\cal J}^{\beta\alpha\mu}_{(1)}(k_1, k_2; Q)\,+\,{\cal J}^{\beta\alpha\mu}_{(2)}(k_1, k_2; Q).\nonumber
\end{eqnarray}
For the computation of ${\cal J}^{\beta\alpha\mu}_{(1)}(k_1, k_2; Q)$ it is sufficient to calculate the trace of the Dirac matrices and integrate over $k$
\begin{eqnarray}\label{label30}
&&{\cal J}^{\beta\alpha,\mu}_{(1)}(k_1, k_2; Q)\,=\,[1\,+\,2\,J_2(M_{\rm N})]\,
[(Q\,+\,2\,k_1\,+\,k_2)^{\alpha}g^{\beta\mu}\,+\nonumber\\
&&+\,(Q\,+\,2\,k_1\,+\,k_2)^{\beta}g^{\mu\alpha}\,+\,(Q\,+\,2\,k_1\,+\,k_2)^{\mu}g^{\alpha\beta}\,-\\
&&-\,2\,(k_1\,+\,k_2)^{\alpha} g^{\beta\mu}\,-\,2\,k^{\beta}_1 g^{\mu\alpha}]\,,\nonumber
\end{eqnarray}
where $J_2(M_{\rm N})$ describes a divergent contribution depending on the cut--off $\Lambda_{\rm D}$. Due to the inequality $\Lambda_{\rm D} \ll M_{\rm N}$ we can neglect $J_2(M_{\rm N})$ with respect to the convergent contribution.

In order to compute ${\cal J}^{\beta\alpha\mu}_{(2)}(k_1, k_2; Q)$ we have to integrate first over $k$ directions. This gives 
\begin{eqnarray}\label{label31}
&&{\cal J}^{\beta\alpha\mu}_{(2)}(k_1, k_2; Q)\,=\nonumber\\
&&=\,\int\frac{d^4k}{\pi^2i}\,\Bigg[\frac{1}{2}\frac{M^2_{\rm N}\,k^2}{(M^2_{\rm N}-k^2)^{\,4}}-\frac{1}{6}\frac{k^4}{(M^2_{\rm N}-k^{\,2})^4}\Bigg]\,(3\,Q\,+\,k_1\,+\,k_2)_{\,\lambda}\times \nonumber\\
&&\quad\quad\times {\rm tr}\,(\gamma^{\beta}\gamma^{\lambda}\gamma^{\alpha}\gamma^{\mu}+\gamma^{\beta}\gamma^{\alpha}\gamma^{\lambda}\gamma^{\mu}+\gamma^{\beta}\gamma^{\alpha}\gamma^{\mu}\gamma^{\lambda})\,=\\
&&=\,-\,\frac{1}{9}[1\,+\,6\,J_2(M_{\rm N})]\,[(3\,Q\,+\,2\,k_1\,+\,k_2)^{\alpha}g^{\beta\mu}\,+\,(3\,Q\,+\,2\,k_1\,+\,k_2)^{\beta} g^{\mu\alpha}\,+\nonumber\\
&&\quad\,+\,(3\,Q\,+\,2\,k_1\,+\,k_2)^{\mu} g^{\alpha\beta}].\nonumber
\end{eqnarray}
Here we have used the integrals

\parbox{11cm}{\begin{eqnarray*}
\int\,\frac{d^4k}{\pi^2i}\,\frac{1}{(M^2_{\rm N}-k^2)^3}&=&\frac{1}{2\,M^2_{\rm N}}\,,\\
\int\,\frac{d^4k}{\pi^2i}\,\frac{1}{(M^2_{\rm N}-k^2)^4}&=&\frac{1}{6\,M^4_{\rm N}}\,.
\end{eqnarray*}} \hfill
\parbox{1cm}{\begin{eqnarray}\label{label32}
\end{eqnarray}}

\noindent Summarizing the contributions, we get
\begin{eqnarray}\label{label33}
&&{\cal J}^{\beta\alpha\mu}(k_1, k_2; Q)\,=\,\frac{2}{3}\,(Q^{\alpha}g^{\beta\mu}\,+\,Q^{\beta}g^{\mu\alpha}\,+\,Q^{\mu}g^{\alpha\beta})\,+\,\frac{8}{9}\,[1\,+\,\frac{3}{2}\,J_2(M_{\rm N})]\times \nonumber\\
&&\times [(2\,k_1\,+\,k_2)^{\alpha}g^{\beta\mu}\,+\,(2\,k_1\,+\,k_2)^{\beta}\,g^{\mu\alpha}\,+\,(2\,k_1\,+\,k_2)^{\mu} g^{\alpha\beta}]\,+\\
&&+\,[1\,+\,2\,J_2(M_{\rm N})][-\,2\,(k_1\,+\,k_2)^{\alpha}g^{\beta\mu}\,-\,2\,k_1^{\beta}g^{\mu\alpha}].\nonumber
\end{eqnarray}
It is seen that the $Q$--dependence coincides with that obtained by the Gertsein--Jackiw method (\ref{label28}). Due to the arbitrariness of $Q$ we can arbsorb by the $Q$--term the terms having the same Lorentz structure. This brings up the r.h.s. of (\ref{label33}) to the form
\begin{eqnarray}\label{label34}
{\cal J}^{\beta\alpha\mu}(k_1, k_2; Q)&=&
\frac{2}{3}\,(Q^{\alpha}g^{\beta\mu}\,+\,Q^{\beta}g^{\mu\alpha}\,+\,Q^{\mu}g^{\alpha\beta})\,+\nonumber\\
&&+\,[-\,2\,(k_1\,+\,k_2)^{\alpha}g^{\beta\mu}\,-\,2\,k_1^{\beta}\,g^{\mu\alpha}]\,.
\end{eqnarray}
Also we have neglected the divergent contribution. This approximation is valid due to the inequality $\Lambda_{\rm D} \ll M_{\rm N}$.

The effective Lagrangian ${\cal L}_{\,\rm Fig.2c}(x)$ defined by (\ref{label34}) reads
\begin{eqnarray}\label{label35}
&&{\cal L}_{\rm Fig.2c}(x)\,=\,i\,e\,\frac{g^2_{\rm V}}{6\,\pi^2}[(3\,-\,a)\,\partial^{\alpha}D^{\dagger}_{\alpha}(x)\,D_{\beta}(x)\,A^{\beta}(x)\,-\nonumber\\
&&-\,(3\,-\,a)\,D^{\dagger}_{\alpha}(x)\,\partial^{\beta}D_{\beta}(x)\,A^{\alpha}(x)\,-\,b\,D^{\dagger}_{\alpha}(x)\,D_{\beta}(x)\,\partial^{\alpha}\,A^{\beta}(x)\,-\nonumber\\
&&-\,(b\,-\,a)\,D^{\dagger}_{\alpha}(x)\,D_{\beta}(x)\,\partial^{\beta}\,A^{\alpha}(x)\,-\\
&&-\,(a\,-\,b)\,\partial^{\beta}\,D^{\dagger}_{\alpha}(x)\,D^{\alpha}(x)\,A_{\beta}(x)\,+\,b\,D^{\dagger}_{\alpha}(x)\,\partial^{\beta}\,D^{\alpha}(x)\,A_{\beta}(x)\,+\nonumber\\
&&+\,3\,D^{\dagger}_{\alpha}(x)\,D_{\beta}(x)\,(\partial^{\alpha}\,A^{\beta}(x)\,-\,\partial^{\beta}\,A^{\alpha}(x))].\nonumber
\end{eqnarray}
Now we can consider $a$ and $b$ as free parameters that can be fixed from the requirement of the gauge invariance of the effective Lagrangian described by the one--nucleon loop diagram in Fig.~2c. 

Due to the constraints $\partial^{\mu}D^{\dagger}_{\mu}(x)\,=\,\partial^{\mu}D_{\mu}(x)\,=\,0$ the corresponding terms in the Lagrangian (\ref{label34}) can be dropped 
\begin{eqnarray}\label{label36}
&&{\cal L}_{\rm Fig.2c}(x)\,=\,i\,e\,\frac{g^2_{\rm V}}{6\,\pi^2}[-\,b\,D^{\dagger}_{\alpha}(x)\,D_{\beta}(x)\,\partial^{\alpha}\,A^{\beta}(x)\,-\nonumber\\
&&-\,(b\,-\,a)\,D^{\dagger}_{\alpha}(x)\,D_{\beta}(x)\,\partial^{\beta}\,A^{\alpha}(x)\,-\\
&&-\,(a\,-\,b)\,\partial^{\beta}\,D^{\dagger}_{\alpha}(x)\,D^{\alpha}(x)\,A_{\beta}(x)\,+\,b\,D^{\alpha}(x)\,\partial^{\beta}\,D^{\alpha}(x)\,A^{\beta}(x)\,+\nonumber\\
&&+\,3\,D^{\dagger}_{\alpha}(x)\,D_{\beta}(x)\,(\partial^{\alpha}\,A^{\beta}(x)\,-\,\partial^{\beta}\,A^{\alpha}(x))].\nonumber
\end{eqnarray}
The subsequent transformations are performed by applying the identity
\begin{eqnarray}\label{label37}
&&D^{\dagger}_{\alpha}(x)\,D_{\beta}(x)\,(\partial^{\alpha}\,A^{\beta}(x)\,-\,\partial^{\beta}\,A^{\alpha}(x))\,=\,\\
&&=\partial^{\beta}\,D^{\dagger}_{\alpha}(x)\,D_{\beta}(x)\,A^{\alpha}(x)\,-\,D^{\dagger}_{\alpha}(x)\,\partial^{\alpha}\,D_{\beta}(x)\,A^{\beta}(x),\nonumber
\end{eqnarray}
where we have dropped the total divergence and the terms proportional to $\partial^{\alpha}\,D^{\dagger}_{\alpha}(x)$ and $\partial^{\beta}\,D_{\beta}(x)$.

Then it is convenient to rewrite the Lagrangian (\ref{label36}) as follows
\begin{eqnarray}\label{label38}
&&{\cal L}_{\rm Fig.2c}(x)\,=\,i\,e\,\frac{g^2_{\rm V}}{6\,\pi^2}[-\,(a\,-\,b)\,D^{\dagger}_{\beta\alpha}(x)\,A^{\beta}(x)\,D^{\alpha}(x)\,+\,b\,D^{\beta\alpha}(x)\,A_{\beta}(x)\,D^{\dagger}_{\,\alpha}(x)\,+\nonumber\\
&&-\,(a\,-\,b)\,\partial_{\alpha}\,D^{\dagger}_{\beta}(x)\,D^{\alpha}(x)\,A_{\beta}(x)\,+\,b\,D^{\dagger}_{\alpha}(x)\,\partial^{\alpha}\,D^{\beta}(x)\,A^{\beta}(x)\,-\\
&&-\,b\,D^{\dagger}_{\alpha}(x)\,D_{\beta}(x)\,\partial^{\alpha}\,A^{\beta}(x)\,-\,(b\,-\,a)\,D^{\dagger}_{\alpha}(x)\,D_{\beta}(x)\,\partial^{\beta}\,A^{\alpha}(x)\,-\nonumber\\
&&+\,3\,D^{\dagger}_{\alpha}(x)\,D_{\beta}(x)\,(\partial^{\alpha}\,A^{\beta}(x)\,-\,\partial^{\beta}\,A^{\alpha}(x))].\nonumber
\end{eqnarray}
Putting $a\,=\,2\,b\,$ we bring up the effective Lagrangian (\ref{label38}) to the irreducible form that contains two parts  defining the renormalization of the electric charge of the deuteron and the gauge invariant interaction coinciding with that given by Corben and Schwinger [14]
\begin{eqnarray}\label{label39}
{\cal L}_{\rm Fig.2c}(x)&=&i\,e\,\frac{g^2_{\rm V}}{6\,\pi^2}[-\,b\,D^{\dagger}_{\beta\alpha}(x)\,A^{\beta}(x)\,D^{\alpha}(x)\,+\,b\,D^{\beta\alpha}(x)\,A_{\beta}(x)\,D^{\dagger}_{\alpha}(x)\,+\nonumber\\
&&\quad\quad\,+\,(2\,b\,+\,3)\,D^{\dagger}_{\alpha}(x)\,D_{\beta}(x)\,F^{\alpha\beta}(x)].
\end{eqnarray}
Now the effective Lagrangian ${\cal L}_{\rm Fig.2c}(x)$ is represented in the irreducible form, and we can proceed to the analysis of gauge invariance.

Putting $b\,=\,0$ we remove the finite contributions to the renormalization constant of the electic charge of the deuteron coming from the one--nucleon  loop diagram in Fig.~2c and get a gauge invariant interaction. As a result the effective gauge invariant interaction reads
\begin{eqnarray}\label{label40}
{\cal L}_{\rm CS}(x)\,=\,i\,e\,\frac{g^2_{\rm V}}{2\pi^2}\,D^{\dagger}_{\mu}(x)\,D_{\nu}(x)\,F^{\mu\nu}(x)\,.
\end{eqnarray}
Thus the one--nucleon  loop diagram in Fig.~2c defines the effective Corben--Schwinger Lagrangian ${\cal L}_{\rm CS}(x)$ describing the interaction of the deuteron with the electromagnetic field. 

We should underscore that the ambiguities connected with the shift of the virtual momentum of the diagram in Fig.~2c are the intrinsic peculiarity of fermion loop diagrams, caused by the computation of such  diagrams within the cut--off regularization [15]. So, one cannot consider the parameters $a$ and $b$ connected to a shift of virtual momentum $Q=a\,k_1 + b\,k_2$ as parameters especially introduced in the model.  The requirement of gauge invariance defines unambiguously the parameters $a$ and $b$ and, correspondingly, the effective Lagrangian ${\cal L}_{\rm CS}(x)$. The choice $a\,=\,2\,b$ is unique, since it brings up the Lagrangian (\ref{label38}) to the irreducible form. The next choice $b = 0$ is also unique, for the effective Lagrangian (\ref{label39}) is given by the irreducible form.

One can show that the contributions of the anomalous magnetic dipole moments of the proton and neutron to the effective Corben--Schwinger Lagrangian are proportional to $J_2(M_{\rm N})$ and due to inequality $\Lambda_{\rm D}\ll M_{\rm N}$ are small compared with that given by Eq.~(\ref{label40}).

\section{The effective Aronson interaction in the relativistic field theory model of the deuteron}
\setcounter{equation}{0}

\hspace{0.2in} In this Section we give the detailed derivation of the effective Aronson Lagrangian defined by the one--nucleon loop diagram in Fig.~2d and describing the effective electromagnetic interactions of the deuteron. The effective Lagrangian described by the diagram in Fig.~2d is given by 
\begin{eqnarray}\label{label41}
&&\int\,d^4x\,{\cal L}_{\rm Fig.2d}(x)\,=\,\int\,d^4x\,\int\,\frac{d^4x_1\,d^4k_1}{(2\pi)^4}\,\frac{d^4x_2\,d^4k_2}{(2\pi)^4}\,D_{\alpha\beta}(x)\,D^{\dagger}_{\mu\,\nu}(x_1)\,A_{\lambda}(x_2)\,\times\nonumber\\
&&\times \, e^{-\,i\,k_1\cdot x_1}\,e^{-\,i\,k_2\cdot x_2}\,e^{\,i\,(k_1\,+\,k_2)\cdot x}\,(-\,e)\,\frac{g^2_{\rm T}}{4\,\pi^2}\,\frac{1}{M^2_{\rm D}}\,{\cal J}^{\alpha\beta\mu\nu\lambda}(k_1, k_2; Q)\,,
\end{eqnarray}
and
\begin{eqnarray}\label{label42}
&&{\cal J}^{\alpha\beta\mu\nu\lambda}(k_1, k_2; Q)\,=\,\int\,\frac{d^4k}{\pi^2\,i}\,\times\\
&&\times {\rm tr}\,\Bigg\{\sigma^{\alpha\beta}\,\frac{1}{M_{\rm N}-\hat{k}-\hat{Q}}\,\sigma^{\mu\nu}\,\frac{1}{M_{\rm N}-\hat{k}-\hat{Q}-\hat{k}_1}\,\gamma^{\lambda}\,\frac{1}{M_{\rm N}-\hat{k}-\hat{Q}-\hat{k}_1-\hat{k}_2}\Bigg\}\,,\nonumber
\end{eqnarray}
where the 4--vector $Q= a\,k_1 + b\,k_2$ is due to the shift of the virtual momentum of the nucleons in the diagram in Fig.~2d. The parameters $a$ and $b$ are free. They are not connected with those having been intriduced for the computation of the diagram in Fig.~2c.

In the r.h.s. of Eq.(\ref{label41}) we have used the fields of the physical deuteron, for the renormalization (\ref{label6}) introduces divergent terms that due to inequality $\Lambda_{\rm D}\ll M_{\rm N}$ are small compared with the convergent ones we are keeping for the computation of the Aronson Lagrangian. The Gertsein--Jackiw ambiguity is given by
\begin{eqnarray}\label{label43}
&&\delta\,{\cal J}^{\alpha\beta\mu\nu\lambda}(k_1, k_2; Q)\,=\,\frac{1}{6}\,{\rm tr}(\sigma^{\alpha\beta}\hat{Q}\,\sigma^{\mu\nu}\gamma^{\lambda})\,.
\end{eqnarray}
Now we should proceed to the computation of ${\cal J}^{\alpha\beta\mu\nu\lambda}(k_1, k_2; Q)$. By analogy with (\ref{label22}) we get
\begin{eqnarray}\label{label44}
&&{\cal J}^{\alpha\beta\mu\nu\lambda}(k_1, k_2; Q)\,=\nonumber\\
&&=\,\int\,\frac{d^4k}{\pi^2\,i}\,{\rm tr}\Bigg\{\sigma^{\alpha\beta}\frac{M_{\rm N}+\hat{k}+\hat{Q}}{M^2_{\rm N}-k^{\,2}}\,\Bigg[1\,+\,\frac{2\,k\cdot Q}{M^2_{\rm N}-k^{\,2}}\Bigg]\,
\sigma^{\mu\nu}\frac{M_{\rm N}+\hat{k}+\hat{Q}+\hat{k}_1}{M^2_{\rm N}-k^2}\,\times \nonumber\\
&&\times\,\Bigg[1\,+\,\frac{2\,k\cdot (Q+k_1)}{M^2_{\rm N}-k^2}\Bigg]\,\gamma^{\lambda}\frac{M_{\rm N}+\hat{k}+\hat{Q}+\hat{k}_1+\hat{k}_2}{M^2_{\rm N}-k^2}\Bigg[1\,+\,\frac{2\,k\cdot (Q+k_1+k_2)}{M^2_{\rm N}-k^2}\Bigg]\Bigg\}\,=\nonumber\\
&&=\,\int\frac{d^4k}{\pi^2i}\,\frac{1}{(M^2_{\rm N}-k^2)^3}{\rm tr}\,\{M^2_{\rm N}[\sigma^{\alpha\,\beta}(\hat{k}+\hat{Q})\sigma^{\mu\nu}\gamma^{\lambda}+\sigma^{\alpha\beta}\sigma^{\mu\nu}(\hat{k}+\hat{Q}+\hat{k}_1)\gamma^{\lambda}+\nonumber\\
&&+\sigma^{\alpha\beta}\sigma^{\mu\nu}\gamma^{\lambda}(\hat{k}+\hat{Q}+\hat{k}_1+\hat{k}_2)]+\sigma^{\alpha\beta}(\hat{k}+\hat{Q})\sigma^{\mu\nu}(\hat{k}+\hat{Q}+\hat{k}_1)\gamma^{\lambda}\times \nonumber\\
&&\times (\hat{k}+\hat{Q}+\hat{k}_1+\hat{k}_2)\}\,
\Bigg[1\,+\,\frac{2\,k\cdot (3Q+2k_1+k_2)}{M^2_{\rm N}-k^2}\Bigg]\,=\\
&&=\,\int\frac{d^4k}{\pi^2\,i}\frac{1}{(M^2_{\rm N}-k^2)^3}{\rm tr}\,\{M^2_{\rm N}[\sigma^{\alpha\beta}\hat{Q}\sigma^{\mu\nu}\gamma^{\lambda}+\sigma^{\alpha\beta}\sigma^{\mu\nu}(\hat{Q}+\hat{k}_1)\gamma^{\lambda}+\nonumber\\  
&&+\sigma^{\alpha\beta}\sigma^{\mu\nu}\gamma^{\lambda}(\hat{Q}+\hat{k}_1+\hat{k}_2)]-\frac{1}{2}k^2 \sigma^{\alpha\beta}\hat{Q}\sigma^{\mu\nu}\gamma^{\lambda}\}\,+\nonumber\\
&&+\,2\,\int\frac{d^4k}{\pi^2i}\frac{1}{(M^2_{\rm N}-k^2)^4}{\rm tr}\,\{\frac{1}{2}M^2_{\rm N}k^2 [\sigma^{\alpha\beta}(3\,\hat{Q}+2\,\hat{k}_1+\hat{k}_2)\sigma^{\mu\nu}\gamma^{\lambda}+\nonumber\\
&&+\sigma^{\alpha\beta}\sigma^{\mu\nu}(3\,\hat{Q}+2\,\hat{k}_1+\hat{k}_2)\gamma^{\lambda}+\sigma^{\alpha\beta}\sigma^{\mu\nu}\gamma^{\lambda}(3\,\hat{Q}+2\,\hat{k}_1+\hat{k}_2)]-\nonumber\\
&&-\frac{1}{6}\,k^4\sigma^{\alpha\beta}(3\,\hat{Q}+2\,\hat{k}_1+\hat{k}_2)\sigma^{\mu\nu}\gamma^{\lambda}\}\,=\,{\cal J}^{\alpha\beta\mu\nu\lambda}_{(1)}(k_1, k_2; Q)\,+\,{\cal J}^{\alpha\beta\mu\nu\lambda}_{(2)}(k_1, k_2; Q)\,.\nonumber
\end{eqnarray}
Integrating over $k$ we obtain
\begin{eqnarray}\label{label45}
&&{\cal J}^{\alpha\beta\mu\nu\lambda}_{(1)}(k_1, k_2; Q)\,=\,\frac{1}{4}\,[1+2\,J_2(M_{\rm N})]\,{\rm tr}(\sigma^{\alpha\beta}\hat{Q}\sigma^{\mu\nu}\gamma^{\lambda})\,+\nonumber\\
&&+\frac{1}{2}\,{\rm tr}\,[\sigma^{\alpha\beta}\sigma^{\mu\nu}(\hat{Q}+\hat{k}_1)\gamma^{\lambda}+\sigma^{\alpha\beta}\sigma^{\mu\nu}\gamma^{\lambda}(\hat{Q}+\hat{k}_1+\hat{k}_2)]\,,
\end{eqnarray}
\begin{eqnarray}\label{label46}
&&{\cal J}^{\alpha\beta\mu\nu\lambda}_{(2)}(k_1, k_2; Q)\,=\,-\frac{1}{6}\,[-\frac{5}{6}+J_2(M_{\rm N})]\,{\rm tr}[\sigma^{\alpha\beta}(3\hat{Q}+2\hat{k}_1+\hat{k}_2)\sigma^{\mu\nu}\gamma^{\lambda}]\,-\nonumber\\
&&-\frac{1}{6}\,{\rm tr}\,[\sigma^{\alpha\beta}(3\hat{Q}+2\hat{k}_1+\hat{k}_2)\sigma^{\mu\nu}\gamma^{\lambda}+\\
&&+\sigma^{\alpha\beta}\sigma^{\mu\nu}(3\hat{Q}+2\hat{k}_1+\hat{k}_2)\gamma^{\lambda}+\sigma^{\alpha\beta}\sigma^{\mu\nu}\gamma^{\lambda}(3\hat{Q}+2\hat{k}_1+\hat{k}_2)].\nonumber
\end{eqnarray}
Now we should summarize the contributions and collect like terms
\begin{eqnarray}\label{label47}
{\cal J}^{\alpha\beta\mu\nu\lambda}(k_1, k_2; Q)&=&\frac{1}{6}\,{\rm tr}[\sigma^{\alpha\beta}(\hat{Q}-\frac{1}{6}(2\,\hat{k}_1+\hat{k}_2))\sigma^{\mu\nu}\gamma^{\lambda}]\,+\\
&+&\frac{1}{6}\,{\rm tr}\,[
\sigma^{\alpha\beta}\sigma^{\mu\nu}(\hat{k}_1-\hat{k}_2)\gamma^{\lambda}]\,+\,\frac{1}{6}\,{\rm tr}\,[\sigma^{\alpha\beta}\sigma^{\mu\nu}\gamma^{\lambda}(\hat{k}_1\,+\,2\,\hat{k}_2)]\,.\nonumber
\end{eqnarray}
It is seen that the $Q$--dependence coincides with that obtained by the Gertsein--Jackiw method. Due to the arbitrariness of $Q$ the vector $(2\,\hat{k}_1+\hat{k}_2)/6$ can removed by the redefinition of $Q$. Thereby we get
\begin{eqnarray}\label{label48}
{\cal J}^{\alpha\beta\mu\nu\lambda}(k_1, k_2; Q)&=&\frac{1}{6}\,{\rm tr}(\sigma^{\alpha\beta}\,\hat{Q}\,\sigma^{\mu\nu}\,\gamma^{\lambda})\,+\\
&+&\frac{1}{6}\,{\rm tr}\,
[\sigma^{\alpha\beta}\,\sigma^{\mu\nu}(\hat{k}_1-\hat{k}_2)\gamma^{\lambda}]\,+\,\frac{1}{6}\,{\rm tr}\,
[\sigma^{\alpha\beta}\sigma^{\mu\nu}\gamma^{\lambda}(\hat{k}_1\,+\,2\,\hat{k}_2)].\nonumber
\end{eqnarray}
After the calculation of the traces of the Dirac matrices we obtain 
${\cal J}^{\alpha\beta\mu\nu\lambda}(k_1, k_2; Q)$ leading to the following effective Lagrangian
\begin{eqnarray}\label{label49}
&&{\cal L}_{\rm Fig.2d}(x)\,=\,(-\,i\,e)\,\frac{g^2_{\rm T}}{4\,\pi^2}\,\frac{1}{M^2_{\rm D}}\nonumber\\
&&\Big[\frac{8}{3}\,a\,\partial_{\lambda}\,D^{\dagger \lambda\nu}(x)\,D_{\nu\mu}(x)\,A^{\mu}(x)+\frac{8}{3}\,a\,D^{\dagger\mu\nu}(x)\,\partial_{\lambda}\,D^{\lambda\nu}(x)\,A_{\mu}(x)+\nonumber\\
&&+\frac{8}{3}\,(b+a)\,D^{\dagger}_{\mu\nu}(x)\,D^{\nu\lambda}(x)\,\partial^{\mu}\,A_{\lambda}(x)+\frac{8}{3}\,(b-a)\,D^{\dagger}_{\mu\nu}(x)\,D^{\nu\lambda}(x)\,\partial_{\lambda}\,A^{\mu}(x)-\\
&&-\frac{16}{3}\,D^{\dagger}_{\mu\nu}(x)\,D^{\nu\lambda}(x)\,\partial^{\mu}\,A_{\lambda}(x)+8\,D^{\dagger}_{\mu\nu}(x)\,D^{\nu\lambda}(x)\,(\partial^{\mu}\,A_{\lambda}(x)-\partial_{\lambda}\,A^{\mu}(x))\Big].\nonumber
\end{eqnarray}
For the derivation of the effective Lagrangian (\ref{label49}) we have used the equation of motion
\begin{equation}\label{label50}
\partial_{\lambda}\,D_{\mu\nu}(x)\,+\,\partial_{\mu}\,D_{\nu\lambda}(x)\,+\,\partial_{\nu}\,D_{\lambda\mu}(x)\,=\,0\,.
\end{equation}
The analogous equation of motion is valid for the conjugated field. By collecting like terms in (\ref{label49}) we get
\begin{eqnarray}\label{label51}
&&{\cal L}_{\rm Fig.2d}(x)\,=\,(-\,i\,e)\,\frac{g^2_{\rm T}}{4\,\pi^2}\,\frac{1}{M^2_{\rm D}}\,\Big[\frac{8}{3}\,(b\,+\,a\,-\,1)\,D^{\dagger}_{\mu\nu}(x)\,D^{\nu\lambda}(x)\,\partial^{\mu}\,A_{\lambda}(x)\,+\nonumber\\
&&+\,\frac{8}{3}\,(b\,-\,a\,-\,3)\,D^{\dagger}_{\mu\nu}(x)\,D^{\nu\lambda}(x)\,\partial_{\lambda}\,A^{\mu}(x)\,+\\
&&+\frac{8}{3}\,a\,\partial_{\lambda}\,D^{\dagger\lambda\nu}(x)\,D_{\nu\mu}(x)\,A^{\mu}(x)\,+\,\frac{8}{3}\,a\,D^{\dagger\mu\nu}(x)\,\partial_{\lambda}\,D^{\lambda\nu}(x)\,A_{\mu}(x)\Big].\nonumber
\end{eqnarray}
The third and the fourth terms can be reduced by applying the equation of motion
\begin{eqnarray}\label{label52}
\partial_{\lambda}\,D^{\lambda\nu}(x)\,=\,-\,M^2_{\rm D}\,D^{\nu}(x)\,
\end{eqnarray}
and analoguous for the conjugated field. Then putting
$b\,+\,a\,-\,1\,=\,-\,b\,+\,a\,+\,3$, we obtain $b\,=\,2$, that brings up the effective Lagrangian (\ref{label51}) to the following irreducible form 
\begin{eqnarray}\label{label53}
&&{\cal L}_{\rm Fig.2d}(x)\,=\nonumber\\
&=&i\,e\,\frac{2\,g^2_{\rm T}}{3\,\pi^2}\,a\,\Big[\,D^{\dagger}_{\mu\nu}(x)\,A^{\mu}(x)\,D^{\nu}(x)\,-\,D^{\mu\nu}(x)\,A_{\mu}(x)\,D^{\dagger}_{\nu}(x)\,\Big]\,+\\
&+&i\,e\,\frac{2\,g^2_{\rm T}}{3\,\pi^2}\,\frac{1}{M^2_{\rm D}}\,(1\,+\,a)\,D^{\dagger}_{\mu\nu}(x)\,D^{\nu\lambda}(x)\,{F_{\,\lambda}}^{\mu}(x)\,.\nonumber
\end{eqnarray}
Putting $a\,=\,0$ we remove the finite contributions to the renormalization constant of the electric charge of the deuteron and gain the gauge invariant interaction
\begin{eqnarray}\label{label54}
&&{\cal L}_{\rm Fig.2d}(x)\,=\,i\,e\,\frac{2\,g^2_{\rm T}}{3\,\pi^2}\,\frac{1}{M^2_{\rm D}}\,D^{\dagger}_{\mu\nu}(x)\,D^{\nu\lambda}(x)\,{F_{\lambda}}^{\mu}(x)\,.
\end{eqnarray}
Due to the irreducibility of the Lagrangian (\ref{label53}) the choice $a\,=0\,$ is unique.

The effective Lagrangian (\ref{label54}) coincides fully with that suggested by Aronson [10]. This means that the one-nucleon loop diagram in Fig.~2d defines unambiguously the effective Lagrangian if one imposes the requirement of the gauge invariance. The same result can be gained if one requires the vanishing of finite contributions to the renormalization constant of the electric charge of the deuteron.

By analogy with the computation of the Lagrangian (\ref{label54}) one can obtain the contribution of the anomalous magnetic dipole moments of the proton and neutron to the effective Aronson interaction
\begin{eqnarray}\label{label55}
&&\delta\,{\cal L}_{\rm Ar}(x)\,=\,-\,i\,e\,(\kappa_{\rm p}-\kappa_{\rm n})\,\Bigg(1-\frac{8}{9}\frac{g_{\rm V}}{g_{\rm T}}\Bigg)\,\frac{3\,g^2_{\rm T}}{4\pi^2}\,\frac{1}{M^2_{\rm D}}\,D^{\dagger}_{\mu\nu}(x)\,D^{\nu\lambda}(x)\,{F_{\lambda}}^{\mu}(x)\,.
\end{eqnarray}
As a result the complete expression of the Aronson effective Lagrangian is given by
\begin{eqnarray}\label{label56}
&&{\cal L}_{\rm Ar}(x)=i\,e\,\frac{2\,g^2_{\rm T}}{3\,\pi^2}\Bigg[1-\frac{9}{8}\,(\kappa_{\rm p}-\kappa_{\rm n})\Bigg(1-\frac{8}{9}\frac{g_{\rm V}}{g_{\rm T}}\Bigg)\Bigg]\,\frac{1}{M^2_{\rm D}}D^{\dagger}_{\mu\nu}(x)D^{\nu\lambda}(x){F_{\lambda}}^{\mu}(x).
\end{eqnarray}
Thus we have shown that in the relativistic field theory model of the deuteron and in one--nucleon  loop approximation one can unambiguously compute effective electromagnetic interactions of the deuteron in terms of anomalous magnetic and electric quadrupole moments. This can be obtained under requirement of gauge invariance to every one--nucleon loop diagram separately. 

\section{Radiative neutron--proton capture}
\setcounter{equation}{0}

\hspace{0.2in}  At low energies, the process of the radiative neutron--proton capture n + p $\to$ D + $\gamma$ comes off via electric and magnetic dipole transitions. In the usual notations ${^{2{\rm S}+1}}{\rm L}_{\rm J}$, the deuteron is a ${^3}{\rm S}_{\,1}$ state, and the possible transitions are
\begin{eqnarray*}
{^3}{\rm S}_1\to {^3}{\rm S}_1\,(\mu)\quad,\quad {^3}{\rm P}_{0,1,2}\to {^3}{\rm S}_1\,(d)\quad,\quad {^1}{\rm S}_0\to {^3}{\rm S}_1\,(\mu)\,,
\end{eqnarray*}
where $\mu\,=$ magnetic dipole and $d\,=$ electric dipole. If the energies are low enough, the nucleons are in the S--wave state and thereby only magnetic dipole transitions are possible.

In the S--wave state the magnetic dipole moment operator acts only on spin variables. This implies that the transition ${^3}{\rm S}_1\to {^3}{\rm S}_1$ is forbidden [16]. Thereby the only allowed transition is ${^1}{\rm S}_0\to {^3}{\rm S}_1$. This means that the anomalous magnetic dipole moments of the proton and neutron should give the main contribution [16].

In the relativistic field theory model of the deuteron the interactions of the deuteron with other fields should be obtained through  one--nucleon  loop diagrams. Therefore, for the computation of the effective Lagrangian of the radiative neutron--proton capture we have first to compute the effective Lagrangian, describing low--energy neutron--proton scattering. Keeping to the one--pion exchange we obtain [1]
\begin{eqnarray}\label{label57}
&&{\cal L}^{\rm np}_{\rm eff}(x)=\frac{g^2_{\pi {\rm NN}}}{4M^2_{\pi}}\Big\{[\bar{p}(x) n^c(x)][\bar{n^c}(x) p(x)]+[\bar{p}(x)\gamma^5 n^c(x)][\bar{n^c}(x)\gamma^5 p(x)]+\nonumber\\ &&+[\bar{p}(x)\gamma_{\mu}\gamma^5 n^c(x)][\bar{n^c}(x)\gamma^{\,\mu}\gamma^5 p(x)]+3[\bar{p}(x)\gamma_{\,\mu}n^c(x)][\bar{n^c}(x)\gamma^{,\mu}p(x)]+\\
&&+\frac{3}{2}[\bar{p}(x) \sigma_{\mu\nu} n^c(x)][\bar{n^c}(x)\sigma^{\mu\nu}p(x)]\,\Big\}.\nonumber
\end{eqnarray}
Only terms $[\bar{p}(x)\gamma^5 n^c(x)][\bar{n^c}(x)\gamma^5 p(x)]$ and $[\bar{p}(x)\gamma_{\mu}\gamma^5 n^c(x)][\bar{n^c}(x)\gamma^{\mu}\gamma^5 p(x)]$ contribute to the S--wave of the neutron--proton scattering in the low--energy limit. Therefore, these terms should effect the ${^1}{\rm S}_{0}\to {^3}{\rm S}_{\,1}$ transition in the radiative neutron--proton capture at low energies. The corresponding one--nucleon  loop diagrams are depicted in Figs.~3 and 4.

First let us consider the contribution of the diagram in Fig.~3a. The corresponding Lagrangian reads
\begin{eqnarray}\label{label58}
&&\int\,d^4x\,{\cal L}_{\rm Fig.3a}(x)\,=\nonumber\\
&&=\int\,d^4x\,\int\,\frac{d^4x_1\,d^4k_1}{(2\pi)^4}\,\frac{d^4x_2\,d^4k_2}{(2\pi)^4}\,[\bar{n^c}(x)\gamma^5 p(x)]\,D^{\dagger}_{\mu}(x_1)\,A_{\nu}(x_2)\,\times\\
&&\times \, e^{-\,i\,k_1\cdot x_1}\,e^{-\,i\,k_2\cdot x_2}\,e^{\,i\,(k_1\,+\,k_2)\cdot x}\,i\,e\,\frac{g^2_{\pi\rm NN}}{M^2_{\pi}}\,\frac{g_{\rm V}}{32\pi^2}\,{\cal J}^{\mu\nu}_{\,5}(k_1, k_2; Q)\,,\nonumber
\end{eqnarray}
where
\begin{eqnarray}\label{label59}
&&{\cal J}^{\mu\nu}_5 (k_1, k_2)\,=\\
&&=\int\,\frac{d^4k}{\pi^2i}{\rm tr}\,\Bigg\{\gamma^{\,5}\,\frac{1}{M_{\rm N}-\hat{k}-\hat{Q}}\,\gamma^{\mu}\,\frac{1}{M_{\rm N}-\hat{k}-\hat{Q}-\hat{k}_1}\,\gamma^{\nu}\,\frac{1}{M_{\rm N}-\hat{k}-\hat{Q}-\hat{k}_1-\hat{k}_2}\Bigg\}\,,\nonumber
\end{eqnarray}
and $Q\,=\,a\,k_1\,+\,b\,k_2$ is an arbitrary shift of virtual momentum. Fortunately, the integral ${\cal J}^{\mu\nu}_5 (k_1, k_2; Q)$ does not depend on the shift of the virtual momentum and can be computed unambiguously
\begin{eqnarray}\label{label60}
{\cal J}^{\mu\nu}_5(k_1, k_2; Q)\,=\,\frac{2i}{M_{\rm N}} \,\varepsilon^{\mu\nu\alpha\beta}\,k_{1\,\alpha}\,k_{2\,\beta}\quad (\varepsilon^{\,0\,1\,2\,3}\,=\,1)\,,
\end{eqnarray}
where we have used the relation ${\rm tr}(\gamma^5\gamma^\mu\gamma^\nu\gamma^\alpha\gamma^\beta)=-4i\varepsilon^{\mu\nu\alpha\beta}$. The structure function (\ref{label60}) leads to the effective Lagrangian
\begin{eqnarray}\label{label61}
{\cal L}_{\rm Fig.3a}(x)\,=\,-\,\frac{e}{2M_{\rm N}}\frac{g^2_{\pi\rm NN}}{M^2_{\pi}}\,\frac{g_{\rm V}}{16\pi^2}\,
D^{\,\dagger}_{\mu\nu}(x) {^\ast}F^{\mu\nu}(x)\,[\bar{n^c}(x)\gamma^{\,5}p(x)]\,,
\end{eqnarray}
where ${^\ast}F^{\mu\nu}(x)\,=\,\frac{1}{2}\varepsilon^{\mu\nu\alpha\beta}F_{\alpha\beta}(x)$. 

Now let us consider the contribution of the diagram in Fig.~3b. This contribution is caused by the anomalous mangnetic moment of the proton. The effective Lagranigan is defined
\begin{eqnarray}\label{label62}
&&\int\,d^4x\,{\cal L}_{\rm Fig.3b}(x)\,=\nonumber\\
&&=\int\,d^4x\,\int\,\frac{d^4x_1\,d^4k_1}{(2\pi)^4}\,\frac{d^4x_2\,d^4k_2}{(2\pi)^4}\,[\bar{n^c}(x)\gamma^5 p(x)]\,D^{\dagger}_{\mu}(x_1)\,F_{\alpha\beta}(x_2)\,\times\\
&&\times \, e^{-\,i\,k_1\cdot x_1}\,e^{-\,i\,k_2\cdot x_2}\,e^{\,i\,(k_1\,+\,k_2)\cdot x}\,e\,\frac{\kappa_{\,\rm p}}{4M_{\rm N}}\frac{g^2_{\pi\rm NN}}{M^2_{\pi}}\,\frac{g_{\rm V}}{32\pi^2}\,{\cal J}^{\mu\alpha\beta}_{\,5}(k_1, k_2; Q)\,,\nonumber
\end{eqnarray}
where
\begin{eqnarray}\label{label63}
&&{\cal J}^{\mu\alpha\beta}_5 (k_1, k_2; Q)\,=\\
&&=\int\,\frac{d^4k}{\pi^2i}{\rm tr}\,\Bigg\{\gamma^5\,\frac{1}{M_{\rm N}-\hat{k}-\hat{Q}}\,\gamma^{\mu}\,\frac{1}{M_{\rm N}-\hat{k}-\hat{Q}-\hat{k}_1}\,\sigma^{\alpha\beta}\,\frac{1}{M_{\rm N}-\hat{k}-\hat{Q}-\hat{k}_1-\hat{k}_2}\Bigg\}\,.\nonumber
\end{eqnarray}
Unfortunately, the integral ${\cal J}^{\mu\alpha\beta}_5 (k_1, k_2; Q)$ depends on the shift of the virtual momentum. Therefore, the leading contribution in the momentum expansion is fully arbitrary and is given
\begin{eqnarray}\label{label64}
{\cal J}^{\mu\alpha\beta}_5 (k_1, k_2; Q)\,=\,-\,2\,i\, \,\varepsilon^{\mu\alpha\beta\nu}\,Q_{\nu}\,=\,-\,2\,i\, \,\varepsilon^{\mu\alpha\beta\nu}\,(a\,k_1\,+\,b\,k_2)_{\nu},
\end{eqnarray}
where $a$ and $b$ are arbitrary parameters. The contribution of the momentum $k_2$ produces in the Lagrangian the operator $\partial_{\,\mu}{^\ast}F^{\,\mu\,\nu}(x)$ that vanishes due to Maxwell$^{\,\prime}\,$s equation of motion, i.e. $\partial_{\,\mu}{^\ast}F^{\,\mu\,\nu}(x)\,=\,0$. Thus only the contribution of the momentum $k_1$ matters. The effective Lagrangian produced by ${\cal J}^{\,\mu\alpha\beta}_{\,5}(k_1, k_2; Q)$ given by (\ref{label64}) reads
\begin{eqnarray}\label{label65}
{\cal L}_{\rm Fig.3b}(x)\,=\,\frac{a}{2}\,\kappa_{\rm p}\,\frac{e}{2M_{\rm N}}\frac{g^2_{\pi\rm NN}}{M^2_{\pi}}\,\frac{g_{\rm V}}{16\pi^2}\,
D^{\dagger}_{\mu\nu}(x) {^\ast}F^{\mu\nu}(x)\,[\bar{n^c}(x)\gamma^5 p(x)]\,.
\end{eqnarray}
The effective Lagrangian corresponding to the diagrams in Figs.~3a is given by
\begin{eqnarray}\label{label66}
{\cal L}_{\rm Fig.3a,b}(x)\,=\,-\,\Bigg(1\,-\,\frac{a}{2}\,\kappa_{\rm p}\Bigg)\,\frac{e}{2M_{\rm N}}\frac{g^2_{\pi\rm NN}}{M^2_{\pi}}\,\frac{g_{\rm V}}{16\pi^2}\,
D^{\,\dagger}_{\mu\nu}(x) {^\ast}F^{\mu\nu}(x)\,[\bar{n^c}(x)\gamma^5 p(x)]\,.
\end{eqnarray}
Now let us discuss how we can fix the parameter $a$. As has been mentioned above the radiative capture n + p $\to$ D + $\gamma$ at low energies is a magnetic dipole transition ${^1}{\rm S}_{0}\to {^3}{\rm S}_{\,1}\,(\mu)$. This implies that the contribution of the proton and neutron to amplitude of the radiative capture n + p $\to$ D + $\gamma$ should be proportional to their magnetic dipole moments. In the case of the proton the total magnetic dipole moment equals $(1+\kappa_{\rm p})$, i.e. $\mu_{\rm p}=1+\kappa_{\rm p}$, whereas the total magnetic dipole moment of the neutron is just its anomalous magnetic dipole moment, i.e. $\kappa_{\rm n}=\mu_{\rm n}$. The effective Lagrangian (\ref{label66}) describes the contribution of the proton, therefore the quantity $(1\,-\,a\,\kappa_{\rm p}/2)$ should be nothing more than the total magnetic dipole moment of the proton $\mu_{\rm p}$. This fixes the arbitrariness of the contribution of the diagram in Fig.~3b. Thus, we have to put $a\,=\,-\,2$. As a result the complete contribution of the diagrams in Figs.~3a, b reads
\begin{eqnarray}\label{label67}
{\cal L}_{\rm Fig.3a,b}(x)\,=\,-\,\mu_{\rm p}\,\frac{e}{2M_{\rm N}}\frac{g^2_{\pi\rm NN}}{M^2_{\pi}}\,\frac{g_{\rm V}}{16\pi^2}\,
D^{\dagger}_{\mu\nu}(x) {^\ast}F^{\mu\nu}(x)\,[\bar{n^c}(x)\gamma^5 p(x)]\,.
\end{eqnarray}
Thus we have fixed the ambiguity introduced by the diagram in Fig.~3b by applying the selection rule ${^1}{\rm S}_{0}\to {^3}{\rm S}_{\,1}\,(\mu)$.

By extending the procedure of the computation of the diagrams in Figs.~3a,b on other diagrams in Fig.~3 we get the following complete effective Lagrangian
\begin{eqnarray}\label{label68}
{\cal L}_{\rm Fig.3}(x)\,=\,-\,(\mu_{\rm p}-\mu_{\rm n})\,\frac{e}{2M_{\rm N}}\frac{g^2_{\pi\rm NN}}{M^2_{\pi}}\,\frac{g_{\rm V}}{16\pi^2}\,
D^{\dagger}_{\mu\nu}(x) {^\ast}F^{\mu\nu}(x)\,[\bar{n^c}(x)\gamma^5 p(x)]\,.
\end{eqnarray}
The contributions of the diagrams describing the interaction of the deuteron with the anomalous magnetic dipole moments of the proton and neutron via the tensor nucleon current, i.e. proportional to the constant $g_{\rm T}$, are divergent and due to the inequality $\Lambda_{\rm D}\ll M_{\rm N}$ are small compared with the convergent ones. Thereby the contributions proportional to the coupling constant $g_{\rm T}$ can be neglected and do not appear in the amplitude of the radiative neutron--proton capture n + p $\to$ D + $\gamma$.

The contribution of the $[\bar{p}(x)\gamma_{\mu} \gamma^5 n^c(x)][\bar{n^c}(x)\gamma^{\mu}\gamma^5 p(x)]$ interaction can be computed by analogy to that given above. As a result we obtain
\begin{eqnarray}\label{label69}
{\cal L}_{\rm Fig.4}(x)\,=\,-\,i\,e\,(\mu_{\rm p}-\mu_{\rm n})\,\frac{g^2_{\pi\rm NN}}{M^2_{\pi}}\,\frac{g_{\rm V}}{32\pi^2}\,
D^{\dagger}_{\mu}(x) {^\ast}F^{\mu\nu}(x)\,[\bar{n^c}(x)\gamma_{\nu}\gamma^5 p(x)].
\end{eqnarray}
The total effective low--energy Lagrangian, describing the radiative neutron--proton capture, is defined by the sum of the effective Lagrangians (\ref{label68}) and (\ref{label69}) and reads

\parbox{11cm}{\begin{eqnarray*}
{\cal L}_{\rm eff}(x)&=&-\,(\mu_{\rm p}-\mu_{\rm n})\,\frac{e}{2M_{\rm N}}\frac{g^2_{\pi\rm NN}}{M^2_{\pi}}\,\frac{g_{\rm V}}{16\pi^2}\,
D^{\dagger}_{\mu\nu}(x) {^\ast}F^{\mu\nu}(x)\,[\bar{n^c}(x)\gamma^5 p(x)]\,+\\
&&-i\,e\,(\mu_{\rm p}-\mu_{\rm n})\,\frac{g^2_{\pi\rm NN}}{M^2_{\pi}}\,\frac{g_{\rm V}}{32\pi^2}\,
D^{\dagger}_{\mu}(x) {^\ast}F^{\mu\nu}(x)\,[\bar{n^c}(x)\gamma_{\nu}\gamma^5 p(x)].
\end{eqnarray*}} \hfill
\parbox{1cm}{\begin{eqnarray}\label{label70}
\end{eqnarray}}

\noindent The amplitude of the radiative neutron--proton capture is given by

\parbox{11cm}{\begin{eqnarray*}
{\cal M}({\rm n}+ {\rm p} \to {\rm D} + \gamma)&=&-(\mu_{\rm p}-\mu_{\rm n})\,\frac{e}{2M_{\rm N}}\frac{g^2_{\pi\rm NN}}{M^2_{\pi}}\,\frac{g_{\rm V}}{16\pi^2}\,\varepsilon^{\alpha\beta\mu\nu}\,k_{\alpha}\,e^{\ast}_{\beta}(k)\,e^{\ast}_{\mu}(Q)\times\\
&&\times [\bar{u^c}(p_1)\,(2\,Q_{\nu}\,-\,M_{\rm N}\,\gamma_{\nu})\gamma^{\,5}u(p_2)]\,,
\end{eqnarray*}} \hfill
\parbox{1cm}{\begin{eqnarray}\label{label71}
\end{eqnarray}}

\noindent where $e^{\ast}_{\beta}(k)$ and $e^{\ast}_{\mu}(Q)$ are the polarization 4--vectors of the photon and the deuteron, respectively, then $\bar{u^c}(p_1)$ and $u(p_2)$ are the bispinorial wave--functions of the neutron and proton normalized by $\bar{u^c}(p_1)u^c(p_1)=-2M_{\rm N}$ and $\bar{u}(p_2)u(p_2)=2M_{\rm N}$.

The cross section $\sigma({\rm n} + {\rm p} \to {\rm D} + \gamma)$ of the radiative neutron--proton capture has been measured for thermal neutrons at laboratory velocities $v/c=7.34\cdot 10^{\,-\,6}$ (the absolute value is $v=2.2\cdot 10^{\,5}\,{\rm cm/sec}$) [17] 
\begin{eqnarray}\label{label72}
\sigma({\rm n} + {\rm p} \to {\rm D} + \gamma)_{\exp}\,=\,(334.2 \pm 0.5)\,{\rm mb}\,.
\end{eqnarray}
The cross section $\sigma({\rm n} + {\rm p} \to {\rm D} + \gamma)$ is given by

\parbox{11cm}{\begin{eqnarray*}
\sigma({\rm n} + {\rm p} \to {\rm D} + \gamma)&=&\frac{1}{v}\frac{1}{4E_1E_2}\int\overline{|{\cal M}({\rm n}+ {\rm p} \to {\rm D} + \gamma)|^2}\times\\
&&\times (2\pi)^4\delta^{(4)}(p_1 + p_2 -Q - k)\frac{d^3k}{(2\pi)^3 2E_{\gamma}}\frac{d^3Q}{(2\pi)^3 2E_{\rm D}}\,,
\end{eqnarray*}} \hfill
\parbox{1cm}{\begin{eqnarray}\label{label73}
\end{eqnarray}}

\noindent where $E_1,E_2,E_{\gamma}$ and $E_{\rm D}$ are the energies of the neutron, proton, photon and the deuteron, respectively, $v$ is the laboratory velocity of the neutron, and $\overline{|{\cal M}({\rm n}+ {\rm p} \to {\rm D} + \gamma)|^2}$ is the squared amplitude averaged over polarizations of the neutron and proton and summed over polarizations of the deuteron and photon
\begin{eqnarray}\label{label74}
&&\overline{|{\cal M}({\rm n}+ {\rm p} \to {\rm D} + \gamma)|^2}=\nonumber\\
&&=(\mu_{\rm p}-\mu_{\rm n})^2\,\frac{e^2}{4M^2_{\rm N}}\Bigg[\frac{g^2_{\pi\rm NN}}{M^2_{\pi}}\Bigg]^2\,\frac{g^2_{\rm V}}{256\pi^4}\,\varepsilon^{\alpha\beta\mu\nu}\,\varepsilon_{\lambda\beta\rho\omega}\,k_{\alpha}k^{\lambda}\Bigg({g_{\mu}}^{\rho} - \frac{Q_{\mu}Q^{\rho}}{M^2_{\rm D}}\Bigg)\,\times \frac{1}{4} \times\nonumber\\
&&\times{\rm tr}\Big[(\hat{p}_1 - M_{\rm N})(2Q_{\nu}-M_{\rm N}\gamma_{\nu})\gamma^5 (\hat{p}_2 + M_{\rm N})(-2Q^{\omega} - M_{\rm N} \gamma^{\omega})\gamma^5\Big]=\\
&&=(\mu_{\rm p}-\mu_{\rm n})^2\,\frac{\alpha}{M^2_{\rm N}}\Bigg[\frac{g^2_{\pi\rm NN}}{M^2_{\pi}}\Bigg]^2\,\frac{g^2_{\rm V}}{256\pi^3}\,\varepsilon^{\alpha\beta\mu\nu}\,\varepsilon_{\lambda\beta\rho\omega}\,k_{\alpha}k^{\lambda}\Bigg({g_{\mu}}^{\rho} - \frac{Q_{\mu}Q^{\rho}}{M^2_{\rm D}}\Bigg)\,\times\nonumber\\
&&\times[4(p_1\cdot p_2 + 2M^2_{\rm N})\,Q_{\nu}Q^{\omega} + M^2_{\rm N}(p_{1\,\nu}p^{\omega}_2 + p^{\omega}_1 p_{2\,\nu})-(p_1\cdot p_2 - M^2_{\rm N})\,g^{\omega}_{\nu}]\,,\nonumber
\end{eqnarray}
where $\alpha=e^2/4\pi=1/137$ is the fine structure constant.

In the low--energy limit when the 3--momenta of the neutron and proton tend to zero we obtain
\begin{eqnarray}\label{label75}
&&\overline{|{\cal M}({\rm n}+ {\rm p} \to {\rm D} + \gamma)|^2}=(\mu_{\rm p}-\mu_{\rm n})^2\,\frac{\alpha\,Q_{\rm D}}{128\pi}\Bigg[\frac{g^2_{\pi\rm NN}}{M^2_{\pi}}\Bigg]^2\,\varepsilon^{\alpha\beta\mu\nu}\,\varepsilon_{\lambda\beta\rho\omega}\,k_{\alpha}k^{\lambda}\times\nonumber\\
&&\times[4(p_1\cdot p_2 + 2M^2_{\rm N})\,Q_{\nu}Q^{\omega} + M^2_{\rm N}(p_{1\,\nu}p^{\omega}_2 + p^{\omega}_1 p_{2\,\nu})]\,=\\
&&=(\mu_{\rm p}-\mu_{\rm n})^2\,\frac{25\alpha\,Q_{\rm D}}{32\pi}\Bigg[\frac{g^2_{\pi\rm NN}}{M^2_{\pi}}\Bigg]^2\,M^4_{\rm N}\varepsilon^2_{\rm D}\,,\nonumber
\end{eqnarray}
where we have used the relation (\ref{label16}).

Thus the cross section of the radiative neutron--proton capture reads
\begin{eqnarray}\label{label76}
\sigma({\rm n} + {\rm p} \to {\rm D} + \gamma)\,=\,\frac{1}{v}\,(\mu_{\rm p}-\mu_{\rm n})^2\,\frac{25\alpha\,Q_{\rm D}}{1024\pi^2}\Bigg[\frac{g^2_{\pi\rm NN}}{M^2_{\pi}}\Bigg]^2\,M_{\rm N}\varepsilon^3_{\rm D}\,.
\end{eqnarray}
Putting $v = 7.34\cdot 10^{\,-\,6}$ we get the following theoretical value of $\sigma({\rm n} + {\rm p} \to {\rm D} + \gamma)$
\begin{eqnarray}\label{label77}
\sigma({\rm n} + {\rm p} \to {\rm D} + \gamma)\,=\,156.4\,{\rm mb}\,.
\end{eqnarray}
It is seen that this theoretical value of the cross section disagrees with the experimental data.

However, it should be stressed that we do not have taken into account the resonance contibution to the ${^1}{\rm S}_{0}$ low--energy neutron--proton scattering [18--20]. The account of the resonance contribution gives the amplitude of the radiative neutron--proton capture in the form [21]

\parbox{11cm}{\begin{eqnarray*}
{\cal M}({\rm n}+ {\rm p} \to {\rm D} + \gamma)&=&-(\mu_{\rm p}-\mu_{\rm n})\,\frac{e}{2M_{\rm N}}\frac{g^2_{\pi\rm NN}}{M^2_{\pi}}\,\frac{g_{\rm V}}{16\pi^2}\,\varepsilon^{\alpha\beta\mu\nu}\,k_{\alpha}\,e^{\ast}_{\beta}(k)\,e^{\ast}_{\mu}(Q)\times\\
&&\times \{\bar{u^c}(p_1)\,[2\,Q_{\nu}\Bigg(1\,-\,\frac{16\pi\, M^2_{\pi}}{g^2_{\pi\rm NN}}\,\frac{a_{\rm S}}{M_{\rm N}}\Bigg) -\,M_{\rm N}\,\gamma_{\nu}]\gamma^{\,5}u(p_2)\}\,,
\end{eqnarray*}} \hfill
\parbox{1cm}{\begin{eqnarray}\label{label78}
\end{eqnarray}}

\noindent where $a_{\rm S}=-23.748\,{\rm fm}$ [4] is the ${^1}{\rm S}_{0}$ neutron--proton scattering length that is fully due to the resonance contribution [18--20].

As a result the cross section of the radiative neutron--proton capture is given by

\parbox{11cm}{\begin{eqnarray*}
&&\sigma({\rm n} + {\rm p} \to {\rm D} + \gamma)\,=\,\frac{1}{v}\,(\mu_{\rm p}-\mu_{\rm n})^{\,2}\,\frac{\alpha\,Q_{\rm D}}{1024\pi^2}\,\Bigg[\frac{g^2_{\pi\rm NN}}{M^2_{\pi}}\Bigg]^{\,2}\,\times\\
&&\times\Bigg[1+8\Bigg(1\,-\,\frac{16\pi\, M^2_{\pi}}{g^2_{\pi\rm NN}}\,\frac{a_{\rm S}}{M_{\rm N}}\Bigg)\Bigg(3\,-\,\frac{16\pi\, M^2_{\pi}}{g^2_{\pi\rm NN}}\,\frac{a_{\rm S}}{M_{\rm N}}\Bigg)\Bigg]\,M_{\rm N}\,\varepsilon^{\,3}_{\rm D}\,=\,308.8\,{\rm mb}\,.
\end{eqnarray*}} \hfill
\parbox{1cm}{\begin{eqnarray}\label{label79}
\end{eqnarray}}

\noindent The theoretical value agrees now with the experimental data within an accuracy better than 8$\%$. The cross section (\ref{label79}) predicted in the relativistic field theory model of the deuteron agrees also well with that given by the potential model [22].

\section{The low--energy p + p $\to$ D + ${\rm e}^+$ + $\nu_{\rm e}$ reaction}
\setcounter{equation}{0}

\hspace{0.2in} The process of the two--proton fusion p + p $\to$ D + ${\rm e}^+$ + $\nu_{\rm e}$ plays an important role for the synthesis of deuterons in stars, where the deuterons are destroyed again by the reaction  p + D $\to$ ${^3}{\rm He}$ + $\gamma$ [23]. In  nuclear physics the process  p + p $\to$ D + ${\rm e}^+$ + $\nu_{\rm e}$ is a Gamow--Teller transition governed by the weak axial--vector nucleon current [24].

In the relativistic field theory model of the deuteron the low--energy process p + p $\to$ D + ${\rm e}^+$ + $\nu_{\rm e}$ is closely connected with low--energy proton--proton scattering, i.e. p + p $\to$ p + p. Low--energy proton--proton scattering differs very much from the low--energy neutron--proton scattering. This is  mostly due to the strong contribution of the Coulomb repulsion [25]. The small relative velocities $v$ in the proton--proton system are suppressed by the factor $\exp(-2\pi\alpha/v)$, where $\alpha=e^2/4\pi=1/137$ is the fine structure constant. The factor $\exp(-2\pi\alpha/v)$ is the Gamow penetration factor [25,26]. For the acquaintance with the earliest computation of the cross section of the p + p $\to$ D + ${\rm e}^+$ + $\nu_{\rm e}$ scattering we refer readers to the paper by Bethe and Critchfield [27] and the book by Rosenfeld [28].

For the computation of the effective Lagrangian describing the low--energy proton--proton scattering we assume the one--pion--exchange approximation
\begin{eqnarray}\label{label80}
{\cal L}^{\rm pp}_{\rm eff}(x)=-\,\frac{g^2_{\pi {\rm NN}}(v)}{2M^2_{\pi}}\,[\bar{p}(x)\gamma^5 p(x)]\,[\bar{p}(x)\gamma^5 p(x)]\,.
\end{eqnarray}
Here we have denoted $g_{\pi {\rm NN}}(v)=g_{\pi {\rm NN}}C(v)$ and $C(v)=\sqrt{2\pi\alpha/v}\exp(-\pi\alpha/v)$ takes into account the Coulomb repulsion between protons at low energies [27,29]. 

Applying the Fierz transformation we bring up the Lagrangian (\ref{label80}) to the form
\begin{eqnarray}\label{label81}
&&{\cal L}^{\rm pp}_{\rm eff}(x)=\,\frac{g^2_{\pi {\rm NN}}(v)}{8 M^2_{\pi}}\,\Big\{[\bar{p}(x) p^c(x)][\bar{p^c}(x)p(x)]+[\bar{p}(x)\gamma^5 p^c (x)][\bar{p^c}(x)\gamma^5 p(x)]+\nonumber\\ &&+[\bar{p}(x)\gamma_{\mu}\gamma^5 p^c(x)][\bar{p^c}(x)\gamma^{\mu}\gamma^5 p(x)]+\frac{1}{2}[\bar{p}(x)\sigma_{\mu\nu}p^c(x)][\bar{p^c}(x)\sigma^{\mu\nu} p(x)]\,\Big\}.
\end{eqnarray}
The process p + p $\to$ D + ${\rm e}^+$ + $\nu_{\rm e}$ should run via the intermediate W--boson exchange, i.e. p + p $\to$ D + W$^+$ $\to$ D + ${\rm e}^+$ + $\nu_{\rm e}$. The Lagrangian describing the electroweak interactions of the W--boson with proton, neutron, positron and neutrino reads [2,19]
\begin{eqnarray}\label{label82}
{\cal L}^{\,\rm W}_{\,\rm int}(x)=-\frac{g_{\rm W}}{2\sqrt{2}}[\bar{p}(x)\gamma^{\mu}(1-g_{\rm A}\gamma^5)n(x)+\bar{\nu}_{\rm e}(x)\gamma^{\mu}(1-\gamma^5)e(x)]\,W_{\mu}(x)
+{\rm h.c.}\;.
\end{eqnarray}
Here $g_{\rm W}$ is the electroweak coupling constant connected with the Fermi constant $G_{\rm F}=1.166\times 10^{-5}\,{\rm GeV}^{-2}$ and the W--boson mass $M_{\rm W}$ by the relation [2,19]
\begin{eqnarray}\label{label83}
\frac{g^2_{\rm W}}{8M^2_{\rm W}}\,=\,\frac{G_{\rm F}}{\sqrt{2}}\,,
\end{eqnarray}
where $g_{\rm A}=1.260 \pm 0.012$ is the axial--vector coupling constant  [4] describing the renormalization of the weak axial--vector hadron current by strong interactions.

Only the interaction
$[\bar{p}(x)\gamma_{\mu}\gamma^5 p^c(x)][\bar{p^c}(x)\gamma^{\mu}\gamma^5 p(x)]$ gives in our approach the main contribution to the amplitude of the transition p + p $\to$ D + W. The corresponding one--nucleon loop diagrams are depicted in Fig.~5.

The effective Lagrangian defined by the diagrams in Fig.~5 is given by
\begin{eqnarray}\label{label84}
&&\int\,d^4x\,{\cal L}_{\rm Fig.5}(x)\,=\nonumber\\
&&=\int\,d^4x\,\int\,\frac{d^4x_1\,d^4k_1}{(2\pi)^4}\,\frac{d^4x_2\,d^4k_2}{(2\pi)^4}\,[\bar{p^c}(x)\gamma_{ \alpha}\gamma^5 p(x)]\,D^{\dagger}_{\mu}(x_1)\,W^{\dagger}_{\nu}(x_2)\,\times\\
&&\times \, e^{-\,i\,k_1\cdot x_1}\,e^{-\,i\,k_2\cdot x_2}\,e^{\,i\,(k_1\,+\,k_2)\cdot x}\,i\,g_{\rm A}\,\frac{g_{\rm W}}{2\sqrt{2}}\,\frac{g^2_{\pi\rm NN}(v)}{M^{\,2}_{\pi}}\,\frac{g_{\rm V}}{64\pi^2}\,\bar{{\cal J}}^{\alpha\mu\nu}(k_1, k_2; Q),\nonumber
\end{eqnarray}
where
\begin{eqnarray}\label{label85}
&&\bar{{\cal J}}^{\alpha\mu\nu}(k_1, k_2;Q)\,=\\
&&=\int\,\frac{d^4k}{\pi^2 i}{\rm tr}\,\Bigg\{\gamma^{\alpha}\gamma^5\frac{1}{M_{\rm N}-\hat{k}-\hat{Q}}\gamma^{\mu}\frac{1}{M_{\rm N}-\hat{k}-\hat{Q}-\hat{k}_1}\gamma^{\nu}\gamma^5\frac{1}{M_{\rm N}-\hat{k}-\hat{Q}-\hat{k}_1-\hat{k}_2}\Bigg\}\,.\nonumber
\end{eqnarray}
The computation of this integral can be reduced to the computation of the integral (\ref{label22}) defining the Corben--Schwinger Lagrangian, i.e.
\begin{eqnarray}\label{label86}
\bar{{\cal J}}^{\alpha\mu\nu}(k_1, k_2;Q)\,=\,{\cal J}^{\alpha\mu\nu}(k_1, k_2;Q)+\tilde{{\cal J}}^{\alpha\mu\nu}(k_1, k_2;Q)\,,
\end{eqnarray}
where we have denoted
\begin{eqnarray}\label{label87}
&&\tilde{{\cal J}}^{\alpha\mu\nu}(k_1, k_2;Q)\,=\\
&&=-2\int\,\frac{d^4k}{\pi^2 i}{\rm tr}\,\Bigg\{\gamma^{\alpha}\frac{1}{M_{\rm N}-\hat{k}-\hat{Q}}\gamma^{\mu}\frac{1}{M_{\rm N}-\hat{k}-\hat{Q}-\hat{k}_1}\gamma^{\nu}\frac{M_{\rm N}}{M^2_{\rm N}-(k+Q+k_1+k_2)^2}\Bigg\}\,.\nonumber
\end{eqnarray}
The integral $\tilde{{\cal J}}^{\alpha\mu\nu}(k_1, k_2;Q)$ does not depend on the shift of virtual momentum, i.e. on $Q$, and in leading order in the long--wavelength expansion is given by
\begin{eqnarray}\label{label88}
&&\tilde{{\cal J}}^{\alpha\mu\nu}(k_1, k_2;Q) = -2\int\,\frac{d^4k}{\pi^2 i}\frac{M_{\rm N}}{M^2_{\rm N}-(k+Q+k_1+k_2)^2}\times\nonumber\\
&&{\rm tr}\,\Bigg\{\gamma^{\alpha}\frac{1}{M_{\rm N}-\hat{k}-\hat{Q}}\gamma^{\mu}\frac{1}{M_{\rm N}-\hat{k}-\hat{Q}-\hat{k}_1}\gamma^{\nu}\Bigg\}\,=\nonumber\\
&&=-2\int\,\frac{d^4k}{\pi^2 i}\frac{M_{\rm N}}{(M^2_{\rm N}-k^2)^3}\Bigg[1 + \frac{2 k\cdot (3Q + 2 k_1 + k_2)}{M^2_{\rm N}-k^2}\Bigg]\times\nonumber\\
&&{\rm tr}\{\gamma^{\alpha}(M_{\rm N} + \hat{k} + \hat{Q})\gamma^{\mu}(M_{\rm N} + \hat{k} + \hat{Q} + \hat{k}_1) \gamma^{\,\nu}\}=\nonumber\\
&&=-\int\,\frac{d^4k}{\pi^2 i}\frac{2 M^2_{\rm N}}{(M^2_{\rm N}-k^2)^3}{\rm tr}\{\gamma^{\alpha}\hat{Q}\gamma^{\mu} \gamma^{\nu} + \gamma^{\alpha}\gamma^{\mu}(\hat{Q} + \hat{k}_1) \gamma^{\nu}\}-\\
&&- \int\,\frac{d^4k}{\pi^2 i}\frac{M^2_{\rm N}k^2}{(M^2_{\rm N}-k^2)^4}{\rm tr}\{\gamma^{\alpha}(3\hat{Q}+2 \hat{k}_1 +\hat{k}_2)\gamma^{\mu} \gamma^{\nu} + \gamma^{\alpha}\gamma^{\mu}(3 \hat{Q} + 2 \hat{k}_1 +\hat{k}_2)\gamma^{\nu}\}=\nonumber\\
&&=- {\rm tr}\{\gamma^{\alpha}\hat{Q}\gamma^{\mu} \gamma^{\nu} + \gamma^{\alpha}\gamma^{\mu}(\hat{Q} + \hat{k}_1) \gamma^{\nu}\}+\nonumber\\
&&+{\rm tr}\{\gamma^{\alpha}(\hat{Q}+ \frac{2}{3} \hat{k}_1 + \frac{1}{3} \hat{k}_2)\gamma^{\mu} \gamma^{\nu} + \gamma^{\,\alpha}\gamma^{\mu}(\hat{Q} + \frac{2}{3} \hat{k}_1 + \frac{1}{3} \hat{k}_2)\gamma^{\nu}\}=\nonumber\\
&&= \frac{1}{3} {\rm tr}\{\gamma^{\alpha}(2\hat{k}_1 + \hat{k}_2)\gamma^{\mu} \gamma^{\nu} + \gamma^{\alpha}\gamma^{\mu}(-\hat{k}_1 + \hat{k}_2)\gamma^{\nu}\}=\nonumber\\
&&= 4 (k^{\alpha}_1 g^{\mu\nu} - k^{\nu}_1 g^{\alpha\mu}) + \frac{4}{3} (k_1 + 2 k_2)^{\mu} g^{\alpha\nu}\,.\nonumber
\end{eqnarray}
Summing up the contributions (\ref{label88}) and (\ref{label34}) we obtain

\parbox{11cm}{\begin{eqnarray*}
&&\bar{{\cal J}}^{\alpha\mu\nu}(k_1, k_2;Q)\,=\,\frac{10}{3}(k^{\alpha}_1 g^{\nu\mu} - k^{\nu}_1 g^{\mu\alpha}) - \frac{2}{3}(k_1 + k_2)^{\alpha} g^{\nu\mu} - \frac{2}{3} k^{\nu}_2 g^{\mu\alpha} +\\
&& + \frac{2}{3}[(Q - k_1 + k_2)^{\mu} g^{\alpha\nu} + (Q - k_1 + k_2)^{\alpha} g^{\nu\mu} + (Q - k_1 + k_2)^{\nu} g^{\mu\alpha}]\,.
\end{eqnarray*}} \hfill
\parbox{1cm}{\begin{eqnarray}\label{label89}
\end{eqnarray}}

\noindent In order to fix ambiguities produced by the shift of the virtual momentum let us note that $k_2$, being the momentum of the W--boson, coincides with the momentum of the leptonic pair and is small. Therefore, we suggest to keep only the terms leading in $k_1$ expansion
\begin{eqnarray}\label{label90}
\bar{{\cal J}}^{\alpha\mu\nu}(k_1, k_2;Q)\,=\,\frac{10}{3}(k^{\alpha}_1 g^{\nu\mu} - k^{\nu}_1 g^{\mu\alpha}) - \frac{2}{3}(2-a)\,k^{\alpha}_1 g^{\nu\mu} + \frac{2}{3}(a-1)\, k^{\nu}_1 g^{\mu\alpha}.
\end{eqnarray}
Here we have dropped the term proportional to $k^{\mu}_1$ giving in the Lagrangian the total divergence of the deuteron field that vanishes: $\partial^{\mu}D^{\dagger}_{\mu}=0$.

The structure function $\bar{{\cal J}}^{\alpha\mu\nu}(k_1, k_2;Q)$,defined by the Eq.(\ref{label90}), would be invariant under gauge transformations of the deuteron field if $a=3/2$, i.e.
\begin{eqnarray}\label{label91}
\bar{{\cal J}}^{\alpha\mu\nu}(k_1, k_2;Q)\,=\,3\,(k^{\alpha}_1 g^{\nu\mu} - k^{\nu}_1 g^{\mu\alpha}).
\end{eqnarray}
The effective Lagrangian defined by the structure function (\ref{label91}) has the following form
\begin{eqnarray}\label{label92}
{\cal L}_{\rm Fig.5}(x)=-\,g_{\rm A}\,\frac{g_{\rm W}}{2\sqrt{2}}\,\frac{g^2_{\pi\rm NN}(v)}{M^2_{\pi}}\,\frac{3g_{\rm V}}{64\pi^2}\,W^{\dagger\mu}(x)\,D^{\dagger}_{\mu\nu}(x) \,[\bar{p^c}(x)\gamma^{\nu}\gamma^5 p(x)].
\end{eqnarray}
In turn the effective Lagrangian describing the low--energy process p + p $\to$ D + ${\rm e}^+$ + $\nu_{\rm e}$ reads
\begin{eqnarray}\label{label93}
{\cal L}_{\rm eff}(x) = g_{\rm A}\,\frac{G_{\rm F}}{\sqrt{2}}\,\frac{g^2_{\pi\rm NN}(v)}{M^2_{\pi}}\,\frac{3g_{\rm V}}{64\pi^2}\,j^{\mu}(x)\,
D^{\dagger}_{\mu\nu}(x)\,[\bar{p^c}(x)\gamma^{\nu}\gamma^5 p(x)],
\end{eqnarray}
where $j_{\,\mu}(x)=\bar{\nu}_{\rm e}(x)\gamma_{\,\mu}(1-\gamma^{\,5})e(x)$ is the leptonic electroweak current.

In the low--energy limit when the 3--momenta of protons tend to zero the effective Lagrangian (\ref{label93}) can be reduced to the expression
\begin{eqnarray}\label{label94}
{\cal L}_{\rm eff}(x) = i\,g_{\rm A}\,M_{\rm N}\frac{G_{\rm F}}{\sqrt{2}}\,\frac{g^2_{\pi\rm NN}(v)}{M^2_{\pi}}\,\frac{3g_{\rm V}}{32\pi^2}\,\,j^{\mu}(x)\,
D^{\dagger}_{\mu}(x)\,[\bar{p^c}(x)\gamma^5 p(x)]\,,
\end{eqnarray}
where we have used the relations

\parbox{11cm}{\begin{eqnarray*}
&&[\bar{p^c}(x)\gamma^{\nu}\gamma^5 p(x)]\to - g^{\nu\,0}[\bar{p^c}(x)\gamma^5 p(x)]\,\\
&&D^{\,\dagger}_{\mu\,0}(x)\to -\,i\,M_{\rm D}D^{\,\dagger}_{\mu}(x)
\end{eqnarray*}} \hfill
\parbox{1cm}{\begin{eqnarray}\label{label95}
\end{eqnarray}}

\noindent that is valid in the limit of low 3--momenta of protons and the deuteron. 

The amplitude defined by the effective Lagrangian (\ref{label94}) reads

\parbox{11cm}{\begin{eqnarray*}
{\cal M}({\rm p} + {\rm p} \to {\rm D} + {\rm e}^+ \nu_{e})&=&i\,C(v)\,g_{\rm A} M_{\rm N} \frac{G_{\rm F}}{\sqrt{2}}\frac{g^2_{\pi\rm NN}}{M^2_{\pi}}\,\frac{3g_{\rm V}}{32\pi^2}\,\times\\
&&e^{\ast}_{\mu}(Q)\,[\bar{u}(k_{\nu})\gamma^{\mu}(1-\gamma^5)v(k_{\rm e})]\,[\bar{u^c}(p_1) \gamma^5 u(p_2)]\,,
\end{eqnarray*}} \hfill
\parbox{1cm}{\begin{eqnarray}\label{label96}
\end{eqnarray}}

\noindent where $\bar{u}(k_{\nu})$ and $v(k_{\rm e})$ are the Dirac bispinors of the neutrino and positron. We have separated the factor $C(v)$, describing the Coulomb repulsion, that is convenient for the sequent calculations.

The cross section of the low--energy p + p $\to$ D + ${\rm e}^+$ + $\nu_{\rm e}$ reaction is defined

\parbox{11cm}{\begin{eqnarray*}
\sigma({\rm p} + {\rm p} \to {\rm D} + {\rm e}^+ + \nu_{\rm e})\,=\,\frac{1}{v}\,\frac{1}{4E_1E_2}\int\,\overline{|{\cal M}({\rm p} + {\rm p} \to {\rm D} + {\rm e}^+ + \nu_{\rm e})|^2}\,\times\\
\times (2\pi)^4\,\delta^{(4)}(p_1+p_2-Q-k_{\rm e}-k_{\nu})\,\frac{d^3Q}{(2\pi)^3 2E_{\rm D}}\frac{d^3k_{\rm e}}{(2\pi)^3 2E_{\rm e}}\frac{d^3k_{\nu}}{(2\pi)^3 2E_{\nu}}\,,
\end{eqnarray*}} \hfill
\parbox{1cm}{\begin{eqnarray}\label{label97}
\end{eqnarray}}

\noindent where $v$ is a relative velocity of protons, and $\overline{|{\cal M}({\rm p} + {\rm p} \to {\rm D} + {\rm e}^+ +\nu_{e})|^2}$ is the squared amplitude averaged over polarizations of protons and summed over polarizations of final particles, i.e.
\begin{eqnarray}\label{label98}
&&\overline{|{\cal M}({\rm p} + {\rm p} \to {\rm D} + {\rm e}^+ \nu_{e})|^2}= C^2(v)\,g^2_{\rm A}M^4_{\rm N}\frac{9G^2_{\rm F}Q_{\rm D}}{2^{10}\pi^2}\Bigg[\frac{g^2_{\pi\rm NN}}{M^2_{\pi}}\Bigg]^2\Bigg(g_{\alpha\beta}-\frac{Q_{\alpha}Q_{\beta}}{M^2_{\rm D}}\Bigg)\times\nonumber\\
&&\times {\rm tr}\{(\hat{k}_{\rm e}-m_{\rm e})\gamma^{\alpha}(1-\gamma^5)\hat{k}_{\nu}\gamma^{\beta}(1-\gamma^5)\} \times \frac{1}{4} \times\\
&&\times \frac{1}{4}\times{\rm tr}\{(\hat{p}_1-M_{\rm N})\gamma^5 (1+\gamma^5\hat{s}_1) (\hat{p}_2+M_{\rm N})\gamma^5(1+\gamma^5\hat{s}_2)\},\nonumber
\end{eqnarray}
where $m_{\rm e}=0.511\;{\rm MeV}$ is the mass of positron, $s_1$ and $s_2$ are 4--vectors of polarization of interacting protons such as $s^2_1 = s^2_2 = -1$ and $s_1\cdot p_1 = s_2\cdot p_2 =0$ [2].

Since protons interact at low energies, we calculate the traces over Dirac matrices in the low--energy limit
\begin{eqnarray}\label{label99}
&&\Bigg(g_{\alpha\beta} -\frac{Q_{\alpha}Q_{\beta}}{M^2_{\rm D}}\Bigg) {\rm tr}\{(\hat{k}_{\rm e}-m_{\rm e})\gamma^{\alpha}(1-\gamma^5)\hat{k}_{\nu}\gamma^{\beta}(1-\gamma^5)\} = \nonumber\\
&&= 2\,\Bigg(g_{\alpha\beta} -\frac{Q_{\alpha}Q_{\beta}}{M^2_{\rm D}}\Bigg) {\rm tr}\{\hat{k}_{\rm e} \gamma^{\alpha} \hat{k}_{\nu} \gamma^{\beta}\} = -24\,\Bigg( E_{\rm e} E_{\nu} - \frac{1}{3}\vec{k}_{\rm e}\cdot \vec{k}_{\nu}\,\Bigg) ,\\
&&\frac{1}{16}\times{\rm tr}\{(\hat{p}_1-M_{\rm N})\gamma^5 (1+\gamma^5\hat{s}_1) (\hat{p}_2+M_{\rm N})\gamma^5(1+\gamma^5\hat{s}_2)\} = - \frac{1}{2}\,M^2_{\rm N}(1-\vec{s}_1 \cdot \vec{s}_2).\nonumber
\end{eqnarray}
Substituting Eq.~(\ref{label99}) in Eq.~(\ref{label98}) we get

\parbox{11cm}{\begin{eqnarray*}
\overline{|{\cal M}({\rm p} + {\rm p} \to {\rm D} + {\rm e}^+ \nu_{e})|^2}&=& C^2(v)\,g^2_{\rm A}M^6_{\rm N}\frac{27G^2_{\rm F}Q_{\rm D}}{2^8\pi^2}\Bigg[\frac{g^2_{\pi\rm NN}}{M^2_{\pi}}\Bigg]^2\times\\
&&\times \Bigg( E_{\rm e} E_{\nu} - \frac{1}{3}\vec{k}_{\rm e}\cdot \vec{k}_{\nu}\,\Bigg)\,(1-\vec{s}_1 \cdot \vec{s}_2).
\end{eqnarray*}} \hfill
\parbox{1cm}{\begin{eqnarray}\label{label100}
\end{eqnarray}}

\noindent Now let us take into account the spinorial properties of the interacting protons. At low energies the two--proton fusion proceeds through an intermediate S--wave state with the total spin of the two protons equal to zero (S=0). The former means that the spins of the interating protons are antiparallel, that gives $\vec{s}_1 \cdot \vec{s}_2 = -1$. However, $\vec{s}_1 \cdot \vec{s}_2 = -1$ can be also realized for the two--proton state with a total spin S=1 and zero projection onto the z--axis. Therefore, the factor $(1-\vec{s}_1 \cdot \vec{s}_2)/2$ sould be expanded as $(1-\vec{s}_1 \cdot \vec{s}_2)/2 = (1/2)_{{\rm S}=0}+ (1/2)_{{\rm S}=1}$. Keeping only the S--wave contribution we arrive at the expression

\parbox{11cm}{\begin{eqnarray*}
\overline{|{\cal M}({\rm p} + {\rm p} \to {\rm D} + {\rm e}^+ \nu_{e})|^2}_{\rm S} &=& C^2(v)\,g^2_{\rm A}M^6_{\rm N}\frac{27G^2_{\rm F}Q_{\rm D}}{2^8\pi^2}\Bigg[\frac{g^2_{\pi\rm NN}}{M^2_{\pi}}\Bigg]^2\times\\
&&\times \Bigg( E_{\rm e} E_{\nu} - \frac{1}{3}\vec{k}_{\rm e}\cdot \vec{k}_{\nu}\,\Bigg)\,.
\end{eqnarray*}} \hfill
\parbox{1cm}{\begin{eqnarray}\label{label101}
\end{eqnarray}}

\noindent Now we should carry out the integration over the phase volume of the final D ${\rm e}^+ \nu_{\rm e}$--state
\begin{eqnarray}\label{label102}
&&\int\frac{d^3Q}{(2\pi)^3 2E_{\rm D}}\frac{d^3k_{\rm e}}{(2\pi)^3 2E_{\rm e}}\frac{d^3k_{\nu}}{(2\pi)^3 2E_{\nu}}\times\nonumber\\
&&\times (2\pi)^4\,\delta^{(4)}(p_1+p_2-Q-k_{\rm e}-k_{\nu})\,\Bigg( E_{\rm e} E_{\nu} - \frac{1}{3}\vec{k}_{\rm e}\cdot \vec{k}_{\nu}\,\Bigg)\,=\\
&&= \frac{1}{32\pi^3 M_{\rm N}}\,\int^{\varepsilon_{\rm D}}_{m_{\rm e}}\sqrt{E^2_{\rm e}-m^2_{\rm e}}E_{\rm e}(\varepsilon_{\rm D} - E_{\rm e})^2\,d E_{\rm e} = \frac{\varepsilon^5_{\rm D}}{960\pi^3 M_{\rm N}}\,f(\xi),\nonumber
\end{eqnarray}
where $\xi=m_{\rm e}/\varepsilon_{\rm D}$ and the function $f(\xi)$ is defined by the integral
\begin{eqnarray}\label{label103}
f(\xi)&=&30\,\int^1_{\xi}\sqrt{x^2 -\xi^2}\,x\,(1-x)^2 dx=\nonumber\\
&=&(1 - \frac{9}{2}\,\xi^2 - 4\,\xi^4)\,\sqrt{1-\xi^2} + \frac{15}{2}\,\xi^4\,{\ell n}\Bigg(\frac{1+\sqrt{1-\xi^2}}{\xi}\Bigg) =\\
&=& 0.776\nonumber
\end{eqnarray}
and normalized to unity at $\xi=0$, i.e. $f(0)=1$.

The cross section of the S--wave low--energy two--proton fusion p + p $\to$ D + ${\rm e}^+$ + $\nu_{\rm e}$ is given by 

\parbox{11cm}{\begin{eqnarray*}
\sigma({\rm p} + {\rm p} \to {\rm D} + {\rm e}^+ + \nu_{\rm e})_{\rm S}&=&
\frac{C^2(v)}{v}\,\frac{9g^2_{\rm A}G^2_{\rm F}Q_{\rm D}}{2^{16}5 \pi^5}\,\Bigg[\frac{g^2_{\pi\rm NN}}{M^2_{\pi}}\Bigg]^2\,\varepsilon^5_{\rm D}\,M^3_{\rm N}f\Bigg(\frac{m_{\rm e}}{\varepsilon_{\rm D}}\Bigg) = \\
&=& 1.88\times 10^{-49}\,\frac{C^2(v)}{v}\,{\rm cm}^2.
\end{eqnarray*}} \hfill
\parbox{1cm}{\begin{eqnarray}\label{label104}
\end{eqnarray}}

\noindent The cross section is calculated in units of $\hbar=c=1$. The appearance of the factor $C^2(v)$ agrees well with the result obtained by Bethe and Critchfield [27].

Now we should average the cross section over relative velocities of the proton--proton system with a Maxwell--Boltzmann distribution [27]

\parbox{11cm}{\begin{eqnarray*}
&&<v\,\sigma({\rm p} + {\rm p} \to {\rm D} + {\rm e}^+ + \nu_{\rm e})_{\rm S}> = \\
&&=1.88\times 10^{-49}\,\frac{1}{2}\,\Bigg(\frac{M_N}{4\pi k {\rm T}}\Bigg)^{3/2}\,\int d^3v\,C^2(v)\,\exp\Bigg(-\frac{M_N}{4 k {\rm T}}\,v^2\Bigg) = \\
&&=1.88\times 10^{-49}\,4\pi^2\alpha\,\Bigg(\frac{M_N}{4\pi k {\rm T}}\Bigg)^{3/2}\,\int^{\infty}_0 dv\,v\,\exp\Bigg(-\frac{M_N}{4 k {\rm T}}\,v^2 - \frac{2\pi\alpha}{v}\Bigg) = \\
&&=1.46\times 10^{-40}\sqrt{\frac{M_N}{k{\rm T}}}\int^{\infty}_0 du\,u\,\exp\Bigg(-u^2-\frac{\pi\alpha}{u}\sqrt{\frac{M_{\rm N}}{k{\rm T}}}\Bigg)\,{\rm cm}^3\,{\rm s}^{-1}\,,
\end{eqnarray*}} \hfill
\parbox{1cm}{\begin{eqnarray}\label{label105}
\end{eqnarray}}

\noindent where $k\,=\,8.62\times 10^{-11}\,{\rm MeV}\cdot {\rm K}^{-1}$ is the Boltzmann constant, and ${\rm T}$ is the temperature. Following [27] and computing the integral by a saddle point approximation  we get

\parbox{11cm}{\begin{eqnarray*}
&&<v\,\sigma({\rm p} + {\rm p} \to {\rm D} + {\rm e}^+ + \nu_{\rm e})_{\rm S} > =1.46\times 10^{-40}\,\times\,\frac{1}{\alpha}\,\times\,\frac{2}{\pi}\times\,\Bigg(\frac{1}{3}\Bigg)^2\times\\
&&\times\,\sqrt{\frac{\pi}{3}}\,\times\,\tau^2\,e^{-\tau}\,{\rm cm}^3\,{\rm s}^{-1} = 1.45 \times 10^{-39}\,\tau^2\,e^{-\tau}\,{\rm cm}^3\,{\rm s}^{-1},
\end{eqnarray*}} \hfill
\parbox{1cm}{\begin{eqnarray}\label{label106}
\end{eqnarray}}

\noindent where $\tau$ is connected with the temperature [27]
\begin{eqnarray}\label{label107}
\tau = 3\Bigg(\frac{\alpha^2\pi^2 M_N}{4 k{\rm T}}\Bigg)^{1/3}.
\end{eqnarray}
The temperature dependence of Eq.~(\ref{label106}) coincides with that obtained by Bethe and Critchfield [27].

Setting ${\rm T} = {\rm T}_{\rm c}={\rm T}_6 = 15.5$, measured in unites of $10^6\,{\rm K}$, where  ${\rm T}_{\rm c}$ is the temperature of the solar core in the Standard Solar model [24], that gives $\tau=13.56$ we get the following estimate
\begin{eqnarray}\label{label108}
<v\,\sigma({\rm p} + {\rm p} \to {\rm D} + {\rm e}^+ + \nu_{\rm e})_{\rm S} >\, =\,3.45\times 10^{-43}\,{\rm cm}^3\,{\rm s}^{-1}\,.
\end{eqnarray}
This magnitude is larger by a factor of 2.9 compared to the potential approach [24] (see also [30]). In order to reconcile our result with the solar luminisoty we should assume that the temperature in the solar core equals ${\rm T}_{\rm c}=13.8$ in $10^6\,{\rm K}$ [31]. As has been remarked in Ref.~[31] the  enhancement of the magnitude of the cross section of the two--proton fusion  leads to a decrease of the temperature in the solar core. This gives a strong suppression of the solar neutrino fluxes. 

The increase of the amplitude of the p + p $\to$ D + W transition found in our approach is connected with the computation of the amplitude in terms of one--nucleon loop diagrams. Indeed, the structure function Eq.~(\ref{label91}) defining the effective Lagrangian of the p + p $\to$ D + W transition is due to the contribution of the anomalous part of the one--nucleon loop diagram of the AAV--kind, i.e. with two axial--vector and one vector vertices [15,32]. One cannot expect that such an anomalous contribution should coincide with the contribution described within potential approach in terms of the overlap integral of the wave--functions of the deuteron and two--protons. The ambiguity of the anomalous contribution has been fixed by the requirement of gauge invariance under the gauge transformations of the deuteron field. It is very similar to the fixing of the ambiguity of the well--known axial Adler--Bell--Jackiw anomaly [2,33].

\section{Conclusion}

\hspace{0.2in} In the present paper we have developed the relativistic field theory model of the deuteron that has been suggested in Ref. [1]. We have given the elaborate computation of the Corben--Schwinger and Aronson effective Lagrangians describing the interactions of the deuteron with electromagnetic fields and defining the anomalous magnetic  and electric quadrupole moments of the deuteron. We have adjusted the model to the calculation of the low--energy processes such as the radiative neutron--proton capture n + p $\to$ D + $\gamma$ for thermal neutrons and the fusion of two protons p + p $\to$ D + ${\rm e}^+$ + $\nu_{\rm e}$. This reaction is very important for stellar nucleosynthesis, where the deuterons are destroyed again by the reaction p + D $\to$ ${^3}{\rm He}$ + $\gamma$ [23]. Unfortunately, our model is far from being induced by the dynamics of QCD and seems like an old--fashion approach. It is based on the Lagrangian Eq.~(\ref{label1}) written with the only guiding principle that the deuteron is a bound state of the proton and neutron with spin one [3,10].

We have shown that the model, supplemented by the dynamics of low--energy neutron--proton and proton--proton scattering, is able to describe reasonably the cross sections of the radiative neutron--proton capture in agreement with both experimental data and predictions of the potential model. For the cross section of p + p $\to$ D + ${\rm e}^+$ + $\nu_{\rm e}$  we have obtained an enhancement by a factor 2.9 compared to the potential approach. This enhancement of the cross section could be of relevance for a suppression of the solar neutrino fluxes [31].

We have found that the coupling constant $g_{\rm T}$ does not contribute to the amplitudes of the processes under consideration. Thereby one can assume that the interaction of the deuteron with the antisymmetric tensor nucleon current is less important for the physics of low--energy interactions of the deuteron than the interaction with a vector nucleon current. If it is true, this gives rise the question concerning the need of keeping a nonzero value for the coupling constant $g_{\rm T}$. What would happen if we would put $g_{\rm T}\,=\,0$? 

Putting $g_{\rm T}\,=\,0$ we face the only problem of the fit of the binding energy of the deuteron by the cut--off $\Lambda_{\rm D}\,=\,45.688\,{\rm MeV}$. Using Eq. (\ref{label18}) at $g_{\rm T}\,=\,0$ we obtain the best fit of the binding energy at $\Lambda_{\rm D}\,=\,64.843\,{\rm MeV}$. Recall that  we have identified  $1/\Lambda_{\rm D}$ with the effective radius of the deuteron [3]. By using the new value of the cut--off we get: $r_{\rm D}\,=\,1/\Lambda_{\rm D}\,=\,3.043\,{\rm fm}$. This value agrees well the average value of the deuteron radius, i.e. $<r>\,=\,3.140\,{\rm fm}$ [34].

Putting then $g_{\rm T}\,=\,0$ we simplify our model reducing the number of free parameters and gain the cut--off $\Lambda_{\rm D}\,=\,64.843\,{\rm MeV}$ that should define the average value of the deuteron radius $r_{\rm D}\,=\,1/\Lambda_{\rm D}\,=\,3.043\,{\rm fm}$.

Thus the number of free parameters introduced for the definition of parameters of the physical deuteron are just three: $g_{\rm V}$, $\Lambda_{\rm D}$ and $\lambda$ and  we could fix them unambiguously.
In this connection we should underscore that additional free paprameters appearing due to the ambiguities connected with the shifts of virtual momenta in the diagrams in Figs.~2--5 are intrinsic peculiarities of one--fermion loop diagrams, caused by the computation of such  diagrams within cut--off regularization [15]. Therefore, one cannot consider these parameters as free parameters especially introduced in the model. Of course, the appearance of these ambiguities can lower, from the first glance, the predictive abilities of the model. However, as we have shown, most of these parameters can be unambiguously fixed by the requirement of gauge invariance.

Finally, we want to estimate the theoretical uncertainty of the relativistic field theory model of the deuteron. Unfortunately, in our model there is not a small parameter, like $\alpha=e^2/4\pi=1/137$,  the fine structure constant in QED. There is not a large parameter like the number of quark colours N in the multi--colour extension of QCD with $SU(3)_c\to SU(N)_c$  that allows to apply  large $N$--expansion in powers of $1/{\rm N}$ to quark--gluon diagrams describing the strong low--energy interactions of hadrons.

Our approach is effective and based on the one--nucleon loop diagram approximation for the description of both the self--interactions of the deuteron and coupling of the deuteron to other particles. By describing the self--interactions of the deuteron at the one--nucleon loop approximation we have fitted all parameters characterising the physical deuteron. In this case the only way to estimate the theoretical uncertainty of the approach is to compare the one--nucleon loop data with the two--nucleon loop corrections. In Ref.~[1] we have calculated the two--nucleon loop correction to the binding energy of the deuteron. Setting $g_{\rm T}=0$, this correction reads [1]
\begin{eqnarray*}
(\delta\,\varepsilon_{\rm D})_{\rm two-loop}=\Lambda_{\rm D}\,\frac{11}{3}\,\frac{g^2_{\pi{\rm NN}}}{4\pi}\,\frac{g^2_{\rm V}}{3\pi^3}\Bigg(\frac{\Lambda_{\rm D}}{M_{\pi}}\Bigg)^2 \Bigg(\frac{\Lambda_{\rm D}}{M_{\rm N}}\Bigg)^3 = 0.36\,{\rm MeV}.
\end{eqnarray*}
The numerical value of the two--nucleon loop correction makes up about 16$\%$ of the binding energy $\varepsilon_{\rm D}=2.273\,{\rm MeV}$ calculated in the one--nucleon loop approximation (see Eq.~(\ref{label70})). Thus the magnitude $\Delta =\pm 16\%$ might be accepted as a theoretical uncertainty of the relativistic field theory model of the deuteron. This agrees well with our prediction for the cross section of the radiative neutron--proton capture. Indeed, the theoretical value $\sigma({\rm n} + {\rm p} \to {\rm D} + \gamma) = 308.8\,{\rm mb}$ reproduces the experimental data $\sigma({\rm n} + {\rm p} \to {\rm D} + \gamma)_{\exp}\,=\,(334.2 \pm 0.5)\,{\rm mb}$ with an accuracy of about 8$\%$. By taking into account the theoretical uncertainty the cross section of the radiative neutron--proton capture should read: $\sigma({\rm n} + {\rm p} \to {\rm D} + \gamma) = (308.8 \pm 49.4)\,{\rm mb}$. With regard to the two--proton fusion cross section, we predict an enhancement of $2.9\pm 0.5$ compared to the potential approach.

We acknowledge fruitful discussions with Prof. G. E. Rutkovsky and Dr. H. Leeb, and Dr. M. Meinhart for discussions concerning effective radius of the deuteron. This work was partially supported  by the Fonds zur F\"orderung der wissenschaftlichen Forschung in \"Osterreich (project P10361--PHY)

\newpage

\section*{\bf References}
\begin{description}
\item{[1]}~A.~N.~Ivanov,~N.~I.~Troitskaya, M.~Faber and ~H.~Oberhummer, Phys.~Lett. {\bf B 361} (1995) 74.
\item{[2]}~C.~Itzykson and J.~-~B.~Zuber, in {\it Quantum Field  Theory} (McDraw--Hill), p.p.~6--14 , 1980;
\item{~~~~}~V.~de~Alfaro,~S.~Fubini,~G.~Furlan and ~C.~Rossetti, in {\it Currents~in~Hadron~Physics} (North Holland), Chapt. 5, 1973;
\item{~~~~}~M.~Gell--Mann and ~M.~Levy,  Nuovo.~Cim.~{\bf 16} (1960) 705;
\item{~~~~}~T.~Hakioglu and ~M.~D.~Scadron,  Phys.~Rev. {\bf D 42} (1991) 941;
\item{~~~~}~A.~N.~Ivanov,~M.~Nagy and ~M.~D.~Scadron,  Phys.~Lett. {\bf B 273} (1991) 137;
\item{~~~~}~R. Delbourgo and M. D. Scadron,  Mod.~Phys.~Lett.~{\bf A 10} (1995) 251.
\item{[3]}~Y.~Nambu and ~G.~Jona--Lasinio, Phys.~Rev.~{\bf 122} (1961) 345; ibid. {\bf 124} (1961) 246.
\item{[4]}~S.~De~Benedetti, in {\it Nuclear Reactions}, John Wiley $\&$ Sons, Inc.: New York--London--Sydney, 1967, p. 46.
\item{[5]}~M.~M. Nagels et al., Nucl. Phys. {\bf B 147} (1979) 253.
\item{[6]}~T.~Eguchi, Phys.~Rev.~{\bf D 14} (1976) 2755;
\item{~~~~}~K.~Kikkawa, Prog.~Theor.~Phys.~{\bf 56} (1976) 947;
\item{~~~~}~H.~Kleinert, in Zi\-chi\-chi A. (ed.): Proc.~of Int.~School
of Subnuclear Physics, (1976) p. 289. 
\item{[7]}~T.~Hatsuda and ~T.~Kumihiro,~Progr.~Theor.~Phys.~{\bf 74},
(1985) 765 ; Phys.~Lett.~{\bf B 198}, (1987) 126;
\item{~~~~}~T.~Kumihiro and ~T.~Hatsuda, Phys.~Lett.~{\bf B 206}, (1988) 385.
\item{[8]}~S.~Klint,~M.~Lutz,~V.~Vogl and ~W.~Weise,
Nucl.~Phys.~{\bf A 516}(1990) 429; 469  and references therein.
\item{[9]}~A.~N.~Ivanov,~M.~Nagy and  N.~I.~Troitskaya,
Int.~J.~Mod.~Phys.~{\bf A 7} (1992)  7305;
\item{~~~~}~A.~N.~Ivanov, Int.~J.~Mod.~Phys.~{\bf A 8} (1993) 853; ~Phys.~Lett.~{\bf B 275} (1992) 450;
\item{~~~~}~A.~N.~Ivanov,~N.~I.~Troitskaya and ~M.~Nagy,
Int.~J.~Mod.~Phys.~{\bf A 8} (1992) 2027; 3425;
\item{~~~~}~A.~N.~Ivanov,~N.~I.~Troitskaya and ~M.~Nagy,
Phys.~Lett.~{\bf B 308} (1993) 111;
\item{~~~~}~A.~N.~Ivanov and ~N.~I.~Troitskaya,
Nuovo.~Cim.~{\bf A 108} (1995)  555;
\item{~~~~}~A.~N.~Ivanov,~N.~I.~Troitskaya,~M.~Faber,~M.~Schaler and ~M.~Nagy, Nuovo Cim. {\bf A 107} (1994) 1667; Phys. Lett. {\bf B 336} (1994) 555.
\item{[10]}~B.~Sakita and ~C.~J.~Goebel, Phys.~Rev.~{\bf 127} (1962)  1787;
\item{~~~~}~B.~Sakita, Phys.~Rev.~{\bf 127} (1962)  1800.
\item{[11]}~H.~Aronson, Phys.~Rev.~{\bf 186} (1969)  1434.
\item{[12]}~J.~D.~Jackson, in {\it Klassische Electrodynamik}, Walter de Gruyter, Berlin--New York, 1983.
\item{[13]}~Particle Data Group,~Phys.~Rev.~{\bf D 50} (1994) 1177, Part 1.
\item{[14]}~H.~C.~Corben and ~J.~Schwinger,~Phys.~Rev.~{\bf 58} (1940)  953.
\item{[15]}~I.~S.~Gertsein and ~R.~Jackiw, Phys.~Rev.~{\bf 181} (1969) 1955;
\item{~~~~}~R.~W.~Brown,~C.~C.~Shih and ~B.~L.~Young,~Phys.~Rev.~{\bf 186} (1969) 1491.
\item{[16]}~A.~Di~Giacomo,~G.~Paffuti and ~P.~Rossi, in {\it Selected Problems of Theoretical Physics (with solutions)}, World Scientific,  Singapore -- New Jersey -- London -- Hong Kong, Problem 22, p. 68.
\item{[17]}~A.~E.~Cox,~A.~R.~Wynchank and ~C.~H.~Collie, Nucl.~Rev.~{\bf 74} (1965) 497 and references therein.
\item{[18]}~L.~R.~B.~Elton, in {\it Intoductory Nuclear Physics}, London:~Pitman, 1959.
\item{[19]}~M.~D.~Scadron, in {\it Advanced in Quantum Theory and its Applications Through Feynman Diagrams}, Springer--Verlag, New York--Heidelberg:  1979, pp. 237--239.
\item{[20]}~G.~Breit and ~E.~Wigner, Phys.~Rev.~{\bf 49} (1936) 519.
\item{[21]}~see ref.[3], p. 156.
\item{[22]}~W.~F.~Hornyak, in {\it Nuclear Structure}, Academic Press, New York--San Francisco--London: 1975, p. 495.
\item{[23]}~G.~B\"orner, in {\it The Early Universe (Facts and Fiction)}, Springer--Verlag, Berlin--Heidelberg--New York--London--Paris--Tokyo, 1988, p.111.
\item{[24]}~C.~E.~Rolfs and ~W.~S. Rodney, in {\it Cauldrons in Cosmos}, The University of Chicago Press, Chicago and London, 1988, pp. 328--338.
\item{[25]}~G.~Gamow, Phys. Rev. {\bf 53} (1938) 595.
\item{[26]}~J.~N.~Bachall, in {\it Neutrino Astrophysics}, Cambridge University Press, Cambridge--New York--New Rochelle--Melbourne--Sydney, 1989, p. 60.
\item{[27]}~H.~A.~Bethe and ~C.~L.~Critchfield, Phys. Rev. {\bf 54} (1939) 248.
\item{[28]}~L.~Rosenfeld, in {\it Nuclear Forces}, North--Holland Publishing Company, Amsterdam, 1948, pp. 155--157.
\item{[29]}~F.~L.~Yost,~J.~A.~Wheeler and ~G.~Breit,~Phys.~Rev. {\bf 49} (1936) 174.
\item{[30]}~M.~Kamionkowski and ~J.~N.~Bahcall, Ap. J. {\bf 359} (1991) 884;
\item{~~~~}~J.~N.~Bahcall and ~M.~H.~Pinsonneault, ~Rev.~Mod.~Phys. {\bf 64} (1992) 885.
\item{[31]}~V.~Castellani,~S.~Degl$^{\prime}$Innocenti,~G.~Fiorentini, ~M.~Lissia and ~B.~Ricci,~Phys.~Rep. {\bf 281} (1997) 309.
\item{[32]}~A.~N.~Ivanov,~Sov.~J.~Nucl.~Phys. {\bf 33} (1981) 904.
\item{[33]}~R.~Jackiw, in {\it Lectures on Current Algebra and its Applications}, Princton Series in Physics, Princton University Press, Princeton--New Jersey, 1972.
\item{[34]}~see ref.[22], p. 147.
\end{description}

\newpage
\section*{\bf Figure Captions}
\begin{itemize}
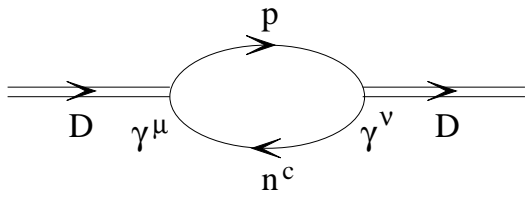
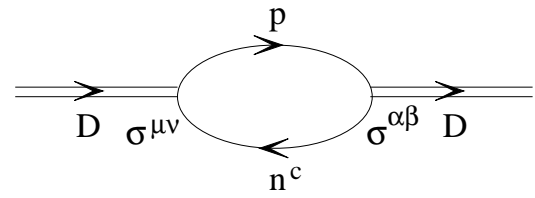
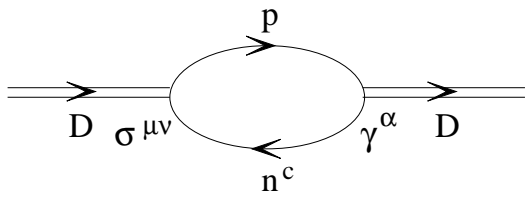
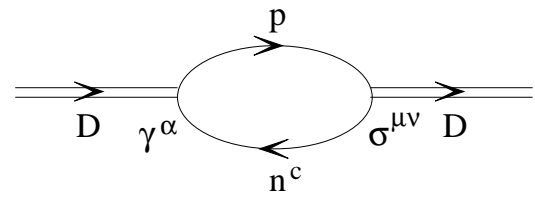
\item{Fig.~1}
One--nucleon loop diagrams contributing to the binding energy of the 
physical deuteron, where $n^c = C\bar{n}^T$ is the field of 
anti--neutron.
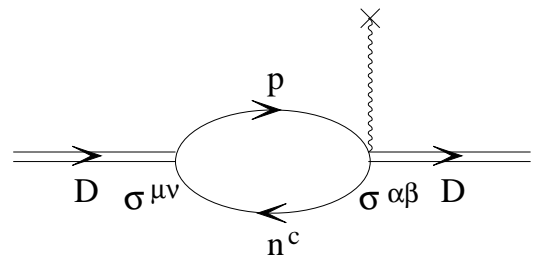
\item{Fig.~2}
One--nucleon loop diagrams describing the effective Corben--Schwinger and Aronson interactions that are responsible for the anomalous magnetic and electric quadrupole moments of the physical deuteron, where $n^c = C\bar{n}^T$ is the field of anti--neutron.
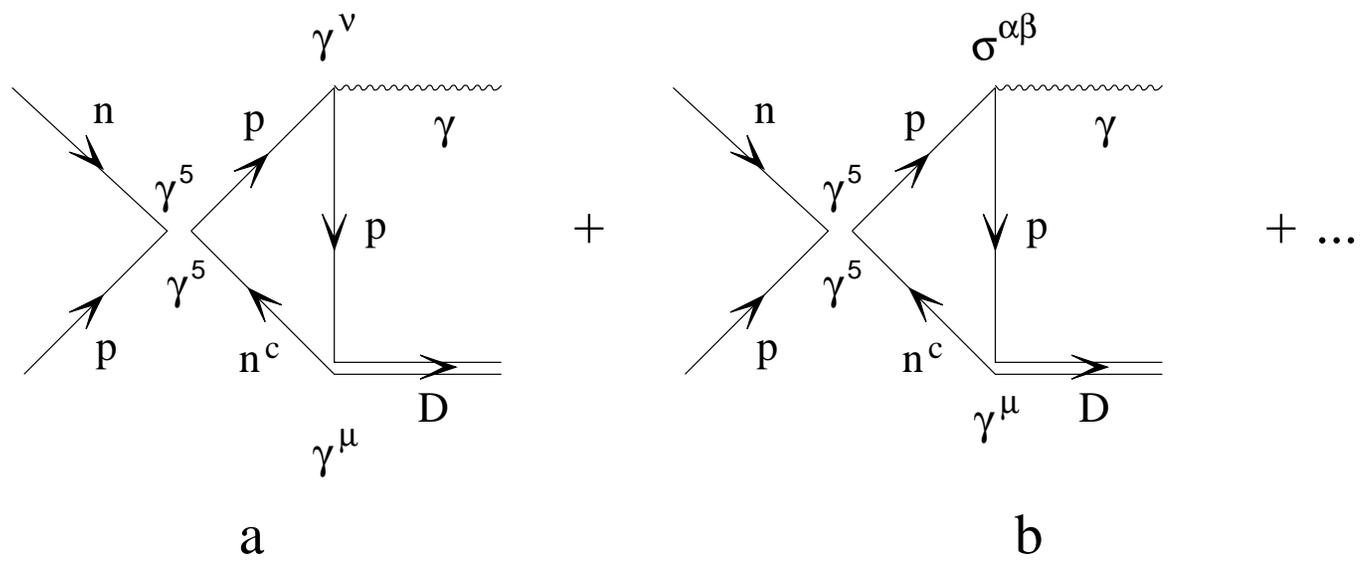
\item{Fig.~3}
The contribution of the $[\bar{p}(x)\gamma^{\,5}n^c(x)][\bar{n^c}(x)\gamma^{\,5}p(x)]$ to the amplitude of the radiative neutron--proton capture.
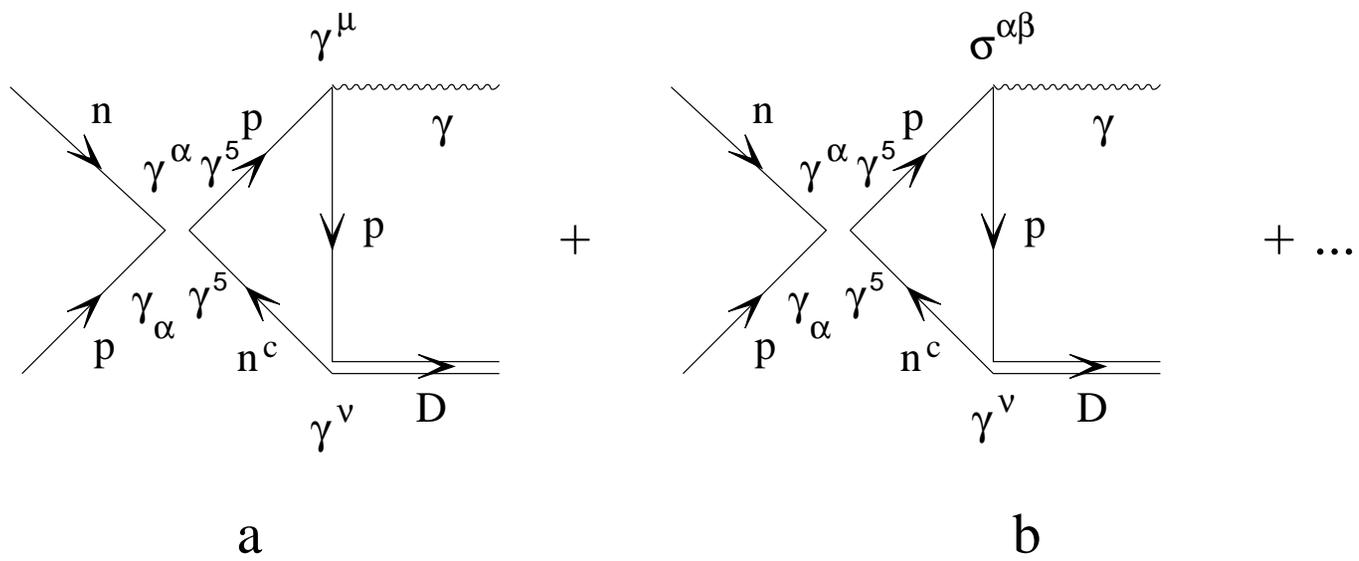
\item{Fig.~4}
The contribution of the $[\bar{p}(x)\gamma_{\,\mu}\gamma^{\,5}n^c(x)][\bar{n^c}(x)\gamma^{\,\mu}\gamma^{\,5}p(x)]$ to the amplitude of the radiative neutron--proton capture.
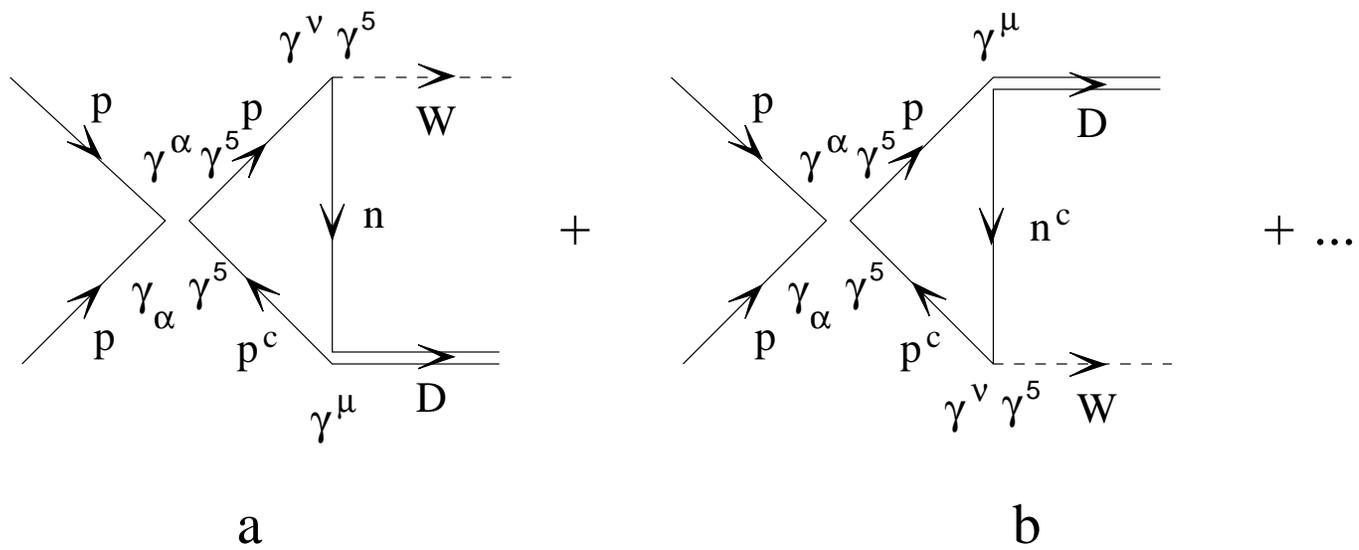
\item{Fig.~5}
The contribution of the $[\bar{p}(x)\gamma_{\,\mu}\gamma^{\,5}p^c(x)][\bar{p^c}(x)\gamma^{\,\mu}\gamma^{\,5}p(x)]$ to the amplitude of the p + p $\to$ D + $e^{\,+}$ + $\nu_{\rm e}$ scattering.
\end{itemize}

\newpage
\section*{Erratum \\ to the paper "On the relativistic field theory model
of the
deuteron II"}
by Ivanov et al. published in Nucl. Phys. A617 (1997) 414.
\vspace{0.2in}

\begin{center}
\begin{abstract}
We correct the value of the cross section for pp--fusion
(p + p $\to$ D + e$^+$ + $\nu_{\rm e}$) calculated in Ref.~[1]. We find a
contribution to the astrophysical factor $\delta S_{\rm pp}(0) = 2.01\times
10^{-25}\,(1 \pm 0.30)\,{\rm MeV}\,{\rm b}$ which is obtained only due to
weak and strong low--energy interactions of the protons and the deuteron
when neglecting the Coulomb repulsion between protons in the intermediate
state. Minor misprints in Ref.~[1] are corrected.
\end{abstract}
\end{center}
\vspace{0.2in}

\noindent {\bf 1.} The enhancement factor 2.9 obtained in Ref.~[1] with
respect to the potential model is mainly related
to the neglect of the electromagnetic mass difference of the proton and the
neutron
by using a value of $W = 2.223$\,MeV instead of 0.932\,MeV in the
phase--space factor
$(W + E)^5f(m_{\rm e}/(W + E))$.
Furthermore, we used a few erroneous combinatorial and dynamical factors
connected with
identical and dynamical properties of the interacting protons.
All of these factors are included now correctly in the Eqs.~1 and 2 given
below.

For the astrophysical factor we give the following new expression
\begin{eqnarray}\label{label1E}
\delta S_{\rm pp}(E) = \alpha\,\frac{9g^2_{\rm A}G^2_{\rm V}Q_{\rm
D}M^4_{\rm N}}
{2560\pi^4}\,\Bigg[\frac{g^2_{\pi\rm NN}}
{4M^2_{\pi}}\Bigg]^2\Bigg(1 - \frac{8\sqrt{2}\pi\,M^2_{\pi}}{g^2_{\rm \pi
NN}}\,
\frac{a_{\rm pp}}{M_{\rm N}}\Bigg)^2\,(W + E)^5\,f\Bigg(\frac{m_{\rm e}}{W
+ E}\Bigg),
\end{eqnarray}
where $E$ is the kinetic energy of the relative movement of the protons,
and the function $f(m_{\rm e}/(W+E))$ is given by Eq.~(6.24) of Ref.~[1].
For $\delta S_{\rm pp}(0)$ we obtain
\begin{eqnarray}\label{label2E}
\delta S_{\rm pp}(0) &=&\alpha\,\frac{9g^2_{\rm A}G^2_{\rm V}Q_{\rm
D}M^4_{\rm N}}{2560\pi^4}\,\Bigg[\frac{g^2_{\pi\rm
NN}}{4M^2_{\pi}}\Bigg]^2\Bigg(1 - \frac{8\sqrt{2}\pi\,M^2_{\pi}}{g^2_{\rm
\pi NN}}\,\frac{a_{\rm pp}}{M_{\rm N}}\Bigg)^2\,W^5\,f\Bigg(\frac{m_{\rm
e}}{W}\Bigg) = \nonumber\\
&=& 2.01\times 10^{-25}\,(1 \pm 0.30){\rm MeV}\,{\rm b},
\end{eqnarray}
where 30$\%$ is the assumed theoretical uncertainty of our model.
The real theoretical uncertainty of the approach can turn out to be much less.
The value $\delta S_{\rm pp}(0) = 2.01\times 10^{-25}\,(1 \pm 0.30){\rm
MeV}\,{\rm b}$
makes up (51.7 $\pm$ 15.5)$\%$ of the value
$S^*_{\rm pp}(0) = 3.89\times 10^{-25}\,(1 \pm 0.011){\rm MeV}\,{\rm b}$
obtained by
Kamionkowski and Bahcall in the potential approach [2].

We stress that the astrophysical factor $\delta S_{\rm pp}$ in
Eqs.~(\ref{label1E})
and (\ref{label2E}) describes only the
contribution of the weak and strong
low--energy interactions and neglects the Coulomb repulsion between protons
in the
one--nucleon loop. The Coulomb repulsion in the initial state is included
in terms of
the Gamow penetration factor $C(v)=\sqrt{2\pi\eta}\exp(-\pi\eta)$ [3], where
$\eta = \alpha/v$, multiplied by the amplitude defined by weak and
strong low--energy interactions of the protons and the deuteron.

Our result concerning the astrophysical factor
$\delta S_{\rm pp}(0) = 2.01\times 10^{-25}\,(1 \pm 0.30){\rm MeV}\,{\rm b}$ is
rather promising. Indeed, as has been shown by Kamionkowski and Bahcall an
important
contribution to the astrophysical factor for pp--fusion comes from the region
where the Coulomb repulsion between the protons
is dominating [2]. Since we have managed to describe about 50$\,\%$ of the
astrophysical
factor obtained in the potential approach keeping only the contribution of
weak and
strong low--energy interactions of the protons and the deuteron, we expect
to get the
contribution of the same order when including the Coulomb repulsion
in the one--nucleon loop. These calculations are
in progress now. We note that we have tested our approach by calculating the
disintegration of the deuteron by reactor antineutrinos
$\bar{\nu}_{\rm e}$ + D $\to$ n + n + e$^+$. This
process is governed by the same physics as the quantity $\delta S_{pp}
(0)$. We find excellent agreement with the potential model results and
the experimental data [4].
\vspace{0.2in}

\noindent {\bf 2.} Eq.~(5.22) of Ref.~[1] should read
\begin{eqnarray*}
{\cal M}({\rm n}+ {\rm p} \to {\rm D} + \gamma)&=&-(\mu_{\rm p}-\mu_{\rm
n})\,\frac{e}{2M_{\rm N}}\frac{g^2_{\pi\rm NN}}{M^2_{\pi}}\,\frac{g_{\rm
V}}{16\pi^2}\,\varepsilon^{\alpha\beta\mu\nu}\,k_{\alpha}\,e^{\ast}_{\beta}(
k)\,e^{\ast}_{\mu}(Q)\times\\
&&\times \{\bar{u^c}(p_1)\,(2\,Q_{\nu} -\,M_{\rm
N}\,\gamma_{\nu})\gamma^{\,5}u(p_2)\}\Bigg(1\,-\,\frac{8\pi\,
M^2_{\pi}}{g^2_{\pi\rm NN}}\,\frac{a_{\rm np}}{M_{\rm N}}\Bigg).
\end{eqnarray*}
Eq.~(5.23) of Ref.~[1] should read
\begin{eqnarray*}
&&\sigma({\rm n} + {\rm p} \to {\rm D} + \gamma)\,=\,\frac{1}{v}\,(\mu_{\rm
p}-\mu_{\rm n})^{\,2}\,\frac{25\alpha\,Q_{\rm
D}}{1024\pi^2}\,\Bigg[\frac{g^2_{\pi\rm
NN}}{M^2_{\pi}}\Bigg]^{\,2}\,\times\\
&&\times\Bigg(1\,-\,\frac{8\pi\, M^2_{\pi}}{g^2_{\pi\rm NN}}\,\frac{a_{\rm
np}}{M_{\rm N}}\Bigg)^2\,M_{\rm N}\,\varepsilon^{\,3}_{\rm D}\,=\,(276\pm
83)\,{\rm mb}\,,
\end{eqnarray*}
where $a_{\rm np}=a_{\rm S}$ in notations of Ref.~[1]. Our result agrees
reasonably well with the experimental data $\sigma({\rm n} + {\rm p} \to
{\rm D} + \gamma)_{\exp}\,=\,(334.2 \pm 0.5)\,{\rm mb}$ [1].

The change of the cross section for radiative capture is caused by the
equal contributions of the pole on the unphysical sheet to the amplitudes
defined by the $\gamma^5\times \gamma^5$ and $\gamma^{\mu}\gamma^5\times
\gamma_{\mu}\gamma^5$ neutron--proton interactions at low energies.

In Ref.~[1] we  have predicted an accuracy of the approach of about 16$\%$.
Since this accuracy has been obtained from the comparison of the
one--nucleon loop and two--nucleon loop contributions to the binding energy
of the deuteron, it should be understood as the accuracy of amplitudes.
Thus,the theoretical accuracy of cross sections should be about 30$\%$.
However, in practice this accuracy can be much better.
\vspace{0.2in}

\noindent {\bf 3.} Minor misprints:

At the end of Chapter 2 of Ref.~[1] the phrase "The contributions of these
diagrams are divergent and can be removed by the renormalization of the
wave-function of the deuteron" should read "The contributions of these
diagrams are divergent and can be removed by the renormalization of the
electric charge of the deuteron".

The phrase "... finite contributions to the renormalization constant of the
wave--function of the deuteron ..."  appearing above Eqs.(3.20), (4.14) and
(4.15) should read "... finite contributions to the renormalization
constant of the electric charge of the deuteron ..."

In the r.h.s of Eq.~(3.4) the common sign instead of $(-)$ should be $(+)$.

In the r.h.s of Eq.(5.18) $g^{\omega}_{\nu}$ should replaced by $M^2_{\rm
N}g^{\omega}_{\nu}$.

Below Eq.~(6.1) $g_{\rm \pi NN}(v) = g_{\rm \pi NN}\,C(v)$ should read
$g^2_{\rm \pi NN}(v) = g^2_{\rm \pi NN}\,C(v)$.

In the r.h.s. of Eq.~(6.23) $\pi^2$ should be replaced by $\pi^3$.

Below Eq.~(6.29) the phrase " This magnitude is smaller ... " should read "
This magnitude is larger ... ".

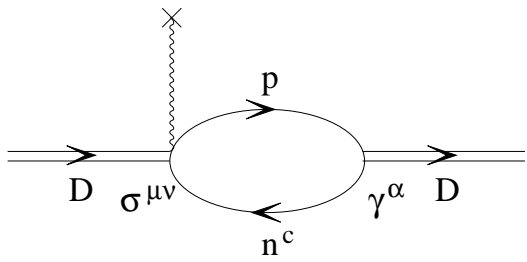
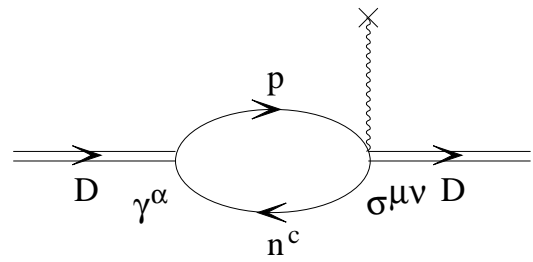
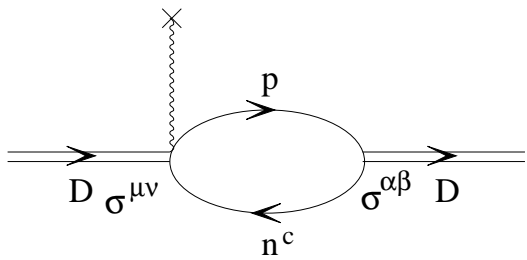
\end{document}